# Sixth-Order Compact Differencing with Staggered Boundary Schemes and 3(2) Bogacki-Shampine Pairs for Pricing Free-Boundary Options

**Chinonso Nwankwo[a,*], Weizhong Dai[b]**

[a] Department of Mathematics and Statistics, University of Calgary, 2500 University Drive NW, Calgary, T2N 1N4, Canada.

[b] Department of Mathematics and Statistics, Louisiana Tech University, Ruston LA 71272, USA
[*] Corresponding author, chinonso.nwankwo@ucalgary.ca, https://orcid.org/0000-0001-5526-1337

## Abstract

We propose a stable sixth-order compact finite difference scheme with a dynamic fifth-order staggered boundary scheme and 3(2) R-K Bogacki and Shampine adaptive time stepping for pricing American style options. To locate, fix and compute the free-boundary simultaneously with option and delta sensitivity, we introduce a Landau transformation. Furthermore, we remove the convective term in the pricing model which could further introduce errors. Hence, an efficient sixth-order compact scheme can easily be implemented. The main challenge in coupling the sixth order compact scheme in discrete form is to efficiently account for the near-boundary scheme. In this work, we introduce novel fifth- and sixth-order Dirichlet near-boundary schemes suitable for solving our model. The optimal exercise boundary and other boundary values are approximated using a high-order analytical approximation obtained from a novel fifth-order staggered boundary scheme. Furthermore, we investigate the smoothness of the first and second derivatives of the optimal exercise boundary which is obtained from this high-order analytical approximation. Coupled with the 3(2) RK-Bogacki and Shampine time integration method, the interior values are then approximated using the sixth order compact operator. The expected convergence rate is obtained, and our present numerical scheme is very fast and gives highly accurate approximations with very coarse grids.

## 1. Model

The system of equations governing the American style put options value and the optimal exercise boundary $s_f(t)$ is given as

$$\frac{\partial P(S,t)}{\partial t}+\frac{1}{2}\sigma^2 S^2\frac{\partial^2 P(S,t)}{\partial S^2}+r\,S\frac{\partial P(S,t)}{\partial S}-rP(S,t)=0, \qquad S>s_f(t); \tag{1a}$$

$$P(S,t)=E-S, \qquad S<s_f(t), \tag{1b}$$

$$P\big(s_f(t),t\big)=E-s_f(t), \qquad \left.\frac{\partial P(S,t)}{\partial S}\right|_{S=s_f(t)}=-1; \tag{1c}$$

$$s_f(0)=E, \qquad P(S,0)=\max(E-S,0). \tag{1d}$$

$$P(\infty,t)=0 \tag{1e}$$

Here, $t=T-t$, $\sigma$ is the volatility, $r$ is the interest rate and $E$ is the strike price.

As shown in (1), the model is a free-boundary problem. There is no closed-form solution for solving this model. Several numerical and semi-analytical methods have been presented for obtaining its approximation. In a finite difference context, it has been modeled as a linear complementary problem, with a penalty method or front-fixing approach. In a linear complementary framework and penalty method, constraints are imposed and it has been demonstrated that the optimal exercise boundary obtained using this approach is not very accurate. In this work, applying the Landau transformation, we solve the model using the front-fixing approach [36]. To this end, let

$$x=\ln S-\ln s_f\,, \qquad P\big(s_f(t)e^x,t\big)=U(x,t). \tag{2}$$

We then obtain a fixed free-boundary PDE problem as follows:

$$\frac{\partial U(x,t)}{\partial t}-\frac{1}{2}\sigma^2\frac{\partial^2 U(x,t)}{\partial x^2}-\left(r+\frac{1}{s_f}\frac{ds_f}{dt}-\frac{\sigma^2}{2}\right)\frac{\partial U(x,t)}{\partial x}+rU(x,t)=0, \qquad x>0. \tag{3}$$

Even though the free-boundary is fixed using the front-fixing approach, as we can see from (3), the transformed model is a nonlinear partial differential equation with a discontinuous coefficient. This is because the derivative of the optimal exercise boundary involved in the coefficient of the convective term is not continuous at payoff. This discontinuity presents a source of non-smoothness in the model. Furthermore, the convective term could further introduce substantial errors when approximating the model. The first challenge will be addressed in the next section. The second challenge is addressed here by further taking the derivative

$$W(x,t) = \frac{\partial U(x,t)}{\partial x}. \tag{4}$$

Hence, we obtain a system of two partial differential equations suitable for implementing an efficient sixth-order compact scheme as follows:

$$\frac{\partial U(x,t)}{\partial t} - \frac{1}{2}\sigma^2 \frac{\partial^2 U(x,t)}{\partial x^2} - \beta_t W(x,t) + rU(x,t) = 0, \qquad x > 0; \tag{5a}$$

$$\frac{\partial W(x,t)}{\partial t} - \frac{1}{2}\sigma^2 \frac{\partial^2 W(x,t)}{\partial x^2} - \beta_t \frac{\partial^2 U(x,t)}{\partial x^2} + rW(x,t) = 0, \qquad x > 0; \tag{5b}$$

$$U(x,t) = E - e^x s_f, \qquad W(x,t) = -e^x s_f, \qquad x \le 0; \tag{5c}$$

$$U(x,0) = 0, \qquad W(x,0) = 0, \qquad x > 0; \tag{5d}$$

$$U(0,t) = E - s_f, \qquad W(0,t) = -s_f, \qquad U(\infty,t) = 0, \qquad W(\infty,t) = 0, \qquad t > 0. \tag{5e}$$

Here, the two systems of PDEs represent the asset options and delta sensitivity, and

$$\beta_t = \left(r + \frac{1}{s_f}\frac{ds_f}{dt} - \frac{\sigma^2}{2}\right). \tag{6}$$

Under sufficient smoothness, a high-order numerical scheme can be used to obtain a more accurate numerical solution with very coarse grids. This feature could be beneficial in saving computational time and improving complexity in high dimensional context. The non-smoothness in the transformed model hampers this possibility [1, 6, 20, 31]. Several authors have implemented a high-order numerical scheme for solving American options using the front-fixing approach. Hajpour and Malek [13] implemented an efficient fifth-order WENO-BDF3 scheme for solving the American options but only recovered a second-order accurate solution. Tangman et al. [31] implemented a fourth-order numerical scheme with coordinate transformation. However, they could not recover the convergence rate that is in good agreement with the theoretical convergence rate. Ballestra [1], using a second-order numerical scheme, recovered a high order convergence rate by implementing a time-variable transformation and Richardson extrapolation. Nwankwo and Dai [23-25] improved the non-smoothness in the transformed model using a high-order analytical approximation. They further implemented a fourth-order compact finite difference scheme and recovered a convergence rate that is in good agreement with the theoretical convergence rate.

Recently, Sari and Gulen [27] implemented a sixth-order finite difference scheme for pricing the American options model using the front-fixing approach. However, they did not address the non-smoothness in the model and approximated the optimal exercise boundary and near-boundary points using the approach of Company et al. [7] which is, at most, second-order accurate in space. This approach could reduce the performance of the sixth-order finite difference scheme. However, Yanbangwai and Moshkin [37] described approaches for improving low-order accurate Dirichlet and Neumann boundary schemes to be consistent with high-order interior schemes using deferred correction techniques. Here, we further acknowledge the recent work of Wang et al. [33] where they implemented a deferred correction method for improving and increasing accuracy in American options using the penalty method.

In this research work, we are particularly interested in the precise computation of the optimal exercise boundary and its derivatives. This is because the left boundary values are not exact. They are rather obtained from the numerical solution of the optimal exercise boundary. Hence, a precise approximation of the optimal exercise boundary will enable us to obtain a more accurate numerical solution of the asset option and delta sensitivity. Furthermore, the coefficient of the convective term in the transformed model involves the derivative of the optimal exercise boundary. It entails a strong need to further obtain a more precise solution of the first derivative of the optimal exercise boundary. The question that remains is at what cost can we achieve these possibilities. For the above purpose, we propose a stable, fast, and very accurate sixth-order compact finite difference scheme with third-order adaptive time stepping based on 3(2) Bogacki and Shampine pairs [2] for pricing American options using a front-fixing approach. Here, we pay more attention to the boundary and near-boundary schemes for approximating the optimal exercise boundary and the boundary values of the asset option and delta sensitivity. We first derive novel fifth- and sixth-order Dirichlet near-boundary schemes suitable for approximating our model. We then construct a fifth-order staggered boundary scheme for approximating the optimal exercise boundary and the boundary scheme. As such, a highly accurate numerical solution can be obtained using a very coarse grid. The rest of the paper is organized as follows. In section 2, we present the proposed sixth-order compact finite difference for solving our model. In section 3, we demonstrate the performance of our proposed method through numerical examples and comparison. We then conclude in section 4.

## 2. Compact Finite Difference

Our computational domain is defined on $[0, x_{max}] \times [0, T]$. $[0, x_{max}]$ is a truncated domain that replaces the semi-positive infinite domain $[0, \infty)$. It has been shown that only negligible error is introduced [7] because the options function vanishes rapidly as $x$ increases. Here, we implement an adaptive time stepping on the time domain. For the space domain, if we denote $x_j$ as a grid point, $h$ as the step size, and $n_x$ as the number of the grid points, then we obtain as follows:

$$x_i = ih, \qquad h = \frac{x_{max}}{n_x}, \qquad i = 0,1,\cdots,n_x. \tag{7}$$

### 2.1. Sixth Order Compact Finite Difference Scheme

We denote the numerical solution of the optimal exercise boundary, asset options, and delta sensitivity as $s_f^n$, $u_i^n$, and $w_i^n$, respectively. Furthermore, in this, we also obtain an approximate solution of the first and second derivatives of the optimal exercise boundary.

To this end, we first present a sixth-order compact finite difference scheme for approximating the option value, delta sensitivity, and optimal exercise boundary simultaneously. As shown in (5), there exists a linear relationship between the optimal exercise boundary and the boundary values of the asset option and delta sensitivity. Hence, we can compute the boundary values from the optimal exercise boundary. Next, we consider $i = 1$. To this end, we introduce new fifth- and sixth-order near-boundary schemes which require some novel ideas as follows:

$$f(x_2,\cdot) = f(x_1,\cdot) + hf'(x_1,\cdot) + \frac{h^2}{2!}f''(x_1,\cdot) + \frac{h^3}{3!}f'''(x_1,\cdot) + \frac{h^4}{4!}f^{(4)}(x_1,\cdot) + \frac{h^5}{5!}f^{(5)}(x_1,\cdot) + \frac{h^6}{6!}f^{(6)}(x_1,\cdot) + \frac{h^7}{7!}f^{(7)}(x_1,\cdot) + O(h^8). \tag{8a}$$

$$f(x_0,\cdot) = f(x_1,\cdot) - hf'(x_1,\cdot) + \frac{h^2}{2!}f''(x_1,\cdot) - \frac{h^3}{3!}f'''(x_1,\cdot) + \frac{h^4}{4!}f^{(4)}(x_1,\cdot) - \frac{h^5}{5!}f^{(5)}(x_1,\cdot) + \frac{h^6}{6!}f^{(6)}(x_1,\cdot) - \frac{h^7}{7!}f^{(7)}(x_1,\cdot) + O(h^8). \tag{8b}$$

If we add $(8a)$ to $(8b)$ and divide by $h^2$, we obtain

$$\frac{f(x_0,\cdot) - 2f(x_1,\cdot) + f(x_2,\cdot)}{h^2} = f''(x_1,\cdot) + \frac{2h^2}{4!}f^{(4)}(x_1,\cdot) + \frac{2h^4}{6!}f^{(6)}(x_1,\cdot) + O(h^6). \tag{9}$$

Here, if we consider a forward finite difference coefficient

$$h^2 f^{(4)}(x_1,\cdot) = 2f''(x_1,\cdot) - 5f''(x_2,\cdot) + 4f''(x_3,\cdot) - f''(x_4,\cdot) + O(h^4), \tag{10}$$

we then obtain

$$14f''(x_1,\cdot) - 5f''(x_2,\cdot) + 4f''(x_3,\cdot) - f''(x_4,\cdot) = \frac{12}{h^2}[\,f(x_0,\cdot) - 2f(x_1,\cdot) + f(x_2,\cdot)] + O(h^4). \tag{11}$$

The fourth-order scheme above is the same as the one presented in the works of Zhao and Zhao et al. [38-40] from which we draw some inspiration to extend beyond fourth-order accuracy. Zhao et al. [38, 40] further presented several fourth-order Dirichlet and Neumann near-boundary schemes suitable for approximating partial differential equations and partial integro differential equations arising in the options model. However, when we approximate our model using the sixth-order combined compact scheme presented in the work of Zhao [39], we observed that the numerical accuracy deteriorates as the step size gets smaller. Notwithstanding, we borrowed insight from their other works [38, 40] to obtain novel fifth- and sixth-order Dirichlet near-boundary schemes. To achieve at least fifth order accuracy near the boundary for $i = 1$, we consider the forward finite difference approximations

$$h^2 f^{(4)}(x_1,\cdot) = \frac{35}{12} f''(x_1,\cdot) - \frac{26}{3} f''(x_2,\cdot) + \frac{19}{2} f''(x_3,\cdot) - \frac{14}{3} f''(x_4,\cdot) + \frac{11}{12} f''(x_5,\cdot) + O(h^5). \tag{12a}$$

$$h^2 f^{(6)}(x_1,\cdot) = f''(x_1,\cdot) - 4f''(x_2,\cdot) + 6f''(x_3,\cdot) - 4f''(x_4,\cdot) + f''(x_5,\cdot) + O(h^5). \tag{12b}$$

Substituting ($12a$) and ($12b$) into (9), we then obtain

$$\frac{897}{60} f''(x_1,\cdot) - \frac{528}{60} f''(x_2,\cdot) + \frac{582}{60} f''(x_3,\cdot) - \frac{288}{60} f''(x_4,\cdot) + \frac{57}{60} f''(x_5,\cdot)$$
$$= \frac{12}{h^2} [\, f(x_0,\cdot) - 2f(x_1,\cdot) + f(x_2,\cdot)] + O(h^5). \tag{13}$$

Furthermore, to achieve sixth-order accuracy near the boundary for $i = 1$, we also consider the forward finite difference approximation

$$h^2 f^{(4)}(x_1,\cdot) = \frac{15}{4} f''(x_1,\cdot) - \frac{77}{6} f''(x_2,\cdot) + \frac{107}{6} f''(x_3,\cdot) - 13f''(x_4,\cdot) + \frac{61}{12} f''(x_5,\cdot) - \frac{5}{6} f''(x_5,\cdot)$$
$$+ O(h^6). \tag{14a}$$

$$h^2 f^{(6)}(x_1,\cdot) = 3f''(x_1,\cdot) - 14f''(x_2,\cdot) + 26f''(x_3,\cdot) - 24f''(x_4,\cdot) + 11f''(x_5,\cdot) - 2f''(x_6,\cdot)$$
$$+ O(h^6). \tag{14b}$$

Substituting ($14a$) and ($14b$) into (9), we then obtain

$$\frac{1902}{120} f''(x_1,\cdot) - \frac{1596}{120} f''(x_2,\cdot) + \frac{2244}{120} f''(x_3,\cdot) - \frac{1656}{120} f''(x_4,\cdot) + \frac{654}{120} f''(x_5,\cdot) - \frac{108}{120} f''(x_6,\cdot)$$
$$= \frac{12}{h^2} [\, f(x_0,\cdot) - 2f(x_1,\cdot) + f(x_2,\cdot)] + O(h^6). \tag{15}$$

Subsequently, for $i = n_x - 1$, we have

$$\frac{897}{60}f''(x_{n_x-1},\cdot) - \frac{528}{60}f''(x_{n_x-2},\cdot) + \frac{582}{60}f''(x_{n_x-3},\cdot) - \frac{288}{60}f''(x_{n_x-4},\cdot) + \frac{57}{60}f''(x_{n_x-5},\cdot)$$
$$= \frac{12}{h^2}\left[ f(x_{n_x-2},\cdot) - 2f(x_{n_x-1},\cdot) + f(x_{n_x},\cdot)\right] + O(h^5). \tag{16a}$$

$$\frac{1902}{120}f''(x_{n_x-1},\cdot) - \frac{1596}{120}f''(x_{n_x-2},\cdot) + \frac{2244}{120}f''(x_{n_x-3},\cdot) - \frac{1656}{120}f''(x_{n_x-4},\cdot) + \frac{654}{120}f''(x_{n_x-5},\cdot)$$
$$- \frac{108}{120}f''(x_{n_x-6},\cdot) = \frac{12}{h^2}\left[ f(x_{n_x-2},\cdot) - 2f(x_{n_x-1},\cdot) + f(x_{n_x},\cdot)\right] + O(h^6). \tag{16b}$$

For $i = 2, \cdots, n_x - 2$, we introduce the well-known sixth-order compact scheme

$$\frac{2}{11}f''(x_{i-1},\cdot) + f''(x_i,\cdot) + \frac{2}{11}f''(x_{i+1},\cdot)$$
$$= \frac{1}{h^2}\left[\frac{3}{44}f(x_{i-2},\cdot) + \frac{12}{11}f(x_{i-1},\cdot) - \frac{51}{22}f(x_i,\cdot) + \frac{12}{11}f(x_{i+1},\cdot) + \frac{3}{44}f(x_{i+2},\cdot)\right] + O(h^6). \tag{17}$$

In matrix-vector form, the discrete system of two equations is presented as follows:

$$A\boldsymbol{u}'' = B_{5,6}\boldsymbol{u} + \boldsymbol{f}_u, \qquad A\boldsymbol{w}'' = B_{5,6}\boldsymbol{w} + \boldsymbol{f}_w. \tag{18}$$

Here, $B_{5,6} \in [B_5, B_6]$ and

$$A = \frac{1}{h^2}\begin{bmatrix} -24 & 12 & 0 & \cdots & & & & 0 \\ \frac{12}{11} & -\frac{51}{22} & \frac{12}{11} & \frac{3}{44} & & & & \vdots \\ \frac{3}{44} & \frac{12}{11} & -\frac{51}{22} & \frac{12}{11} & \frac{3}{44} & & & \\ & \frac{3}{44} & \frac{12}{11} & -\frac{51}{22} & \frac{12}{11} & \frac{3}{44} & & \\ 0 & & \ddots & \ddots & \ddots & \ddots & \ddots & 0 \\ & & & \frac{3}{44} & \frac{12}{11} & -\frac{51}{22} & \frac{12}{11} & \frac{3}{44} \\ & & & & \frac{3}{44} & \frac{12}{11} & -\frac{51}{22} & \frac{12}{11} \\ 0 & & & & \cdots & 0 & 12 & -24 \end{bmatrix}_{n_x-1\times n_x-1}, \tag{19a}$$

$$\boldsymbol{f}_u = \frac{1}{h^2}\begin{bmatrix} 12u_0^2 \\ \frac{3u_0^n}{44} \\ 0 \\ \vdots \\ 0 \\ \frac{3u_{n_x}^n}{44} \\ 12u_{n_x}^n \end{bmatrix}_{n_x-1\times 1}, \quad \boldsymbol{f}_w = \frac{1}{h^2}\begin{bmatrix} 12w_0^2 \\ \frac{3w_0^n}{44} \\ 0 \\ \vdots \\ 0 \\ \frac{3w_{n_x}^n}{44} \\ 12w_{n_x}^n \end{bmatrix}_{n_x-1\times 1}; \tag{19b}$$

$$B_5 = \begin{bmatrix} \frac{897}{60} & -\frac{528}{60} & \frac{582}{60} & -\frac{288}{60} & \frac{57}{60} & 0 & \cdots & 0 \\ \frac{2}{11} & 1 & \frac{2}{11} & & & & & \vdots \\ \vdots & \frac{2}{11} & 1 & \frac{2}{11} & & & & \\ & & \frac{2}{11} & 1 & \frac{2}{11} & & & \\ 0 & & \ddots & \ddots & \ddots & & & 0 \\ \vdots & & & \frac{2}{11} & 1 & \frac{2}{11} & & \\ 0 & \cdots & 0 & \frac{57}{60} & -\frac{288}{60} & \frac{582}{60} & -\frac{528}{60} & \frac{897}{60} \end{bmatrix}_{n_x-1\times n_x-1}, \tag{19c}$$

$$B_6 = \begin{bmatrix} \frac{1902}{120} & -\frac{1596}{120} & \frac{2244}{120} & -\frac{1656}{120} & \frac{654}{120} & -\frac{108}{120} & 0 & \cdots & 0 \\ \frac{2}{11} & 1 & \frac{2}{11} & & & & & & \vdots \\ \vdots & \frac{2}{11} & 1 & \frac{2}{11} & & & & & \\ & & \frac{2}{11} & 1 & \frac{2}{11} & & & & \\ 0 & & \ddots & \ddots & \ddots & & & & 0 \\ \vdots & & & & \frac{2}{11} & 1 & \frac{2}{11} & & \\ 0 & \cdots & 0 & -\frac{108}{120} & \frac{654}{120} & -\frac{1656}{120} & \frac{2244}{120} & -\frac{1596}{120} & \frac{1902}{120} \end{bmatrix}_{n_x-1\times n_x-1}. \tag{19d}$$

We would love to emphasize that the fourth-, fifth- and sixth-order Dirichlet near-boundary schemes we derived above are more suitable for our numerical approximation when we compared them with the one in the work of Mehra [22]. Furthermore, it could easily be seen that we have avoided $f''(x_0,\cdot)$ in the presented schemes. This is a very important feature to observe because of the behavior of the derivative of the option value at the payoff. When we use the sixth order Dirichlet near-boundary scheme in the work of Mehra [22] for approximating our model, we obtained numerical divergence. It is on this notion that our interest was spurred to derive new fifth- and sixth-order Dirichlet near-boundary schemes suitable for our discrete system. Further observation reveals that $A$ is dependent on $h$ while the entries of the matrix $B_5$ is constant. However, $A$ is diagonally dominant.

## 2.2. Dynamic and Fifth Order Staggered Exercise Boundary Scheme

Here, we dynamically construct a fifth-order staggered boundary scheme which will allow us to manipulate how we use grid points arbitrarily near the left boundary when approximating the optimal exercise boundary and the boundary values of the asset option and delta sensitivity. To this end, we first introduce the following Taylor series expansion about the left boundary as follows:

$$f(h_1,\cdot) = f(0,\cdot) + a_1 f'(0,\cdot) + a_2 f''(0,\cdot) + a_3 f'''(0,\cdot) + a_4 f^4(0,\cdot) + a_5 f^5(0,\cdot) + a_6 f^6(0,\cdot) + a_7 f^7(0,\cdot) + O(h^8). \qquad (20a)$$

$$f(h_2,\cdot) = f(0,\cdot) + b_1 f'(0,\cdot) + b_2 f''(0,\cdot) + b_3 f'''(0,\cdot) + b_4 f^4(0,\cdot) + b_5 f^5(0,\cdot) + b_6 f^6(0,\cdot) + b_7 f^7(0,\cdot) + O(h^8). \qquad (20b)$$

$$f(h_3,\cdot) = f(0,\cdot) + c_1 f'(0,\cdot) + c_2 f''(0,\cdot) + c_3 f'''(0,\cdot) + c_4 f^4(0,\cdot) + c_5 f^5(0,\cdot) + c_6 f^6(0,\cdot) + c_7 f^7(0,\cdot) + O(h^8). \qquad (20c)$$

$$f(h_4,\cdot) = f(0,\cdot) + d_1 f'(0,\cdot) + d_2 f''(0,\cdot) + d_3 f'''(0,\cdot) + d_4 f^4(0,\cdot) + d_5 f^5(0,\cdot) + d_6 f^6(0,\cdot) + d_7 f^7(0,\cdot) + O(h^8). \qquad (20d)$$

$$f(h_5,\cdot) = f(0,\cdot) + e_1 f'(0,\cdot) + e_2 f''(0,\cdot) + e_3 f'''(0,\cdot) + e_4 f^4(0,\cdot) + e_5 f^5(0,\cdot) + e_6 f^6(0,\cdot) + e_7 f^7(0,\cdot) + O(h^8). \qquad (20e)$$

$$a_i = \frac{\gamma_1^i h^i}{i!}, \qquad b_i = \frac{\gamma_2^i h^i}{i!}, \qquad c_i = \frac{\gamma_3^i h^i}{i!}, \qquad d_i = \frac{\gamma_4^i h^i}{i!}, \qquad e_i = \frac{\gamma_5^i h^i}{i!}. \qquad (21a)$$

Here,

$$h_1 = \gamma_1 h, \qquad h_2 = \gamma_2 h, \qquad h_3 = \gamma_3 h, \qquad h_4 = \gamma_4 h, \qquad h_5 = \gamma_5 h. \qquad (21b)$$

$\gamma_1,\ \gamma_2,\ \gamma_3,\ \gamma_4,$ and $\gamma_5$ are arbitrary and can be varied to control the distribution of the grid points when approximating the optimal exercise boundary. We leverage this feature to manipulate the distribution of grid points about the left boundary when computing the boundary values, hence presenting a high-order dynamic and staggered boundary scheme. Furthermore, let

$$i_1 = \frac{b_7}{a_7}, \qquad j_1 = \frac{c_7}{b_7}, \qquad k_1 = \frac{d_7}{c_7}, \qquad l_1 = \frac{e_7}{d_7}. \qquad (21c)$$

If we multiply ($20a$) by $i_1$ and subtract it from ($20b$), we obtain as follows:

$$i_1 f(h_1,\cdot) - f(h_2,\cdot) = (i_1 - 1) f(0,\cdot) + (a_1 i_1 - b_1) f'(0,\cdot) + (a_2 i_1 - b_2) f''(0,\cdot) + (a_3 i_1 - b_3) f'''(0,\cdot) + (a_4 i_1 - b_4) f^4(0,\cdot) + (a_5 i_1 - b_5) f^5(0,\cdot) + (a_6 i_1 - b_6) f^6(0,\cdot) + O(h^8). \qquad (22a)$$

Similarly, we can further obtain equations for other differences as follows:

$$j_1 f(h_2,\cdot) - f(h_3,\cdot) = (i_1 - 1) f(0,\cdot) + (b_1 i_1 - c_1) f'(0,\cdot) + (b_2 i_1 - c_2) f''(0,\cdot) + (b_3 i_1 - c_3) f'''(0,\cdot) + (b_4 i_1 - c_4) f^4(0,\cdot) + (b_5 i_1 - c_5) f^5(0,\cdot) + (b_6 i_1 - c_6) f^6(0,\cdot) + O(h^8)\,. \qquad (22b)$$

$$k_1 f(h_3,\cdot) - f(h_4,\cdot)$$
$$= (j_1 - 1)f(0,\cdot) + (c_1 j_1 - d_1)f'(0,\cdot) + (c_2 j_1 - d_2)f''(0,\cdot) + (c_3 j_! - d_3)f'''(0,\cdot) + (c_4 i - d_4)f^4(0,\cdot) + (j_1 c_5 - d_5)f^5(0,\cdot) + (c_6 j_1 - d_6)f^6(0,\cdot) + O(h^8). \quad (22c)$$

$$l_1 f(h_4,\cdot) - f(h_5,\cdot)$$
$$= (l_1 - 1)f(0,\cdot) + (d_1 l_1 - e_1)f'(0,\cdot) + (d_2 l_1 - e_2)f''(0,\cdot) + (d_3 l_1 - e_3)f'''(0,\cdot) + (d_4 l_1 - e_4)f^4(0,\cdot) + (d_5 l_1 - e_5)f^5(0,\cdot) + (d_6 l_1 - e_6)f^6(0,\cdot) + O(h^8). \quad (22d)$$

Continuing similarly, the following dynamic weights and a fifth-order staggered boundary scheme are then obtained

$$-\frac{s_{0,1}}{h^2}f(0,\cdot) + \frac{s_{1,1}}{h^2}f(h_1,\cdot) - \frac{s_{2,1}}{h^2}f(h_2,\cdot) + \frac{s_{3,1}}{h^2}f(h_3,\cdot) - \frac{s_{4,1}}{h^2}f(h_4,\cdot) + f(h_5,\cdot)$$
$$= \frac{v_{1,1}}{h}f'(0,\cdot) + v_{2,1}f''(0,\cdot) + v_{3,1}hf'''(0,\cdot) + O(h^5), \quad (23a)$$

with

$$s_{0,1} = i_4(i_3[i_2(\gamma_1 i_1 - \gamma_2) - (\gamma_2 j_1 - \gamma_3)] - [j_2(\gamma_2 j_1 - \gamma_3) - (\gamma_3 k_1 - \gamma_4)]) - (j_3[j_2(\gamma_2 j_1 - \gamma_3) - (\gamma_3 k_1 - \gamma_4)] - [k_2(\gamma_3 k_1 - \gamma_4) - (\gamma_4 l_1 - \gamma_5)]), \quad (23b)$$

$$s_{1,1} = i_4 i_3 i_2 i_1, \qquad s_{2,1} = i_4 i_3 i_2 + i_4 i_3 j_1 + i_4 j_2 j_1 + j_3 j_2 j_1; \quad (23c)$$

$$s_{3,1} = i_4 i_3 + i_4 j_2 + i_4 k_1 + j_3 k_1 + j_3 j_2 + k_2 k_1, \qquad s_{4,1} = i_4 + j_3 + k_2 + l_1, \qquad s_{5,1} = 1; \quad (23d)$$

$$v_{1,1} = i_4(i_3[i_2(\gamma_1 i_1 - \gamma_2) - (\gamma_2 j_1 - \gamma_3)] - [j_2(\gamma_2 j_1 - \gamma_3) - (\gamma_3 k_1 - \gamma_4)]) - (j_3[j_2(\gamma_2 j_1 - \gamma_3) - (\gamma_3 k_1 - \gamma_4)] - [k_2(\gamma_3 k_1 - \gamma_4) - (\gamma_4 l_1 - \gamma_5)]). \quad (23e)$$

$$v_{2,1} = \frac{1}{2!}i_4(i_3[i_2(\gamma_1^2 i_1 - \gamma_2^2) - (\gamma_2^2 j_1 - \gamma_3^2)] - [j_2(\gamma_2^2 j_1 - \gamma_3^2) - (\gamma_3^2 k_1 - \gamma_4^2)]) - \frac{1}{2!}(j_3[j_2(\gamma_2^2 j_1 - \gamma_3^2) - (\gamma_3^2 k_1 - \gamma_4^2)] - [k_2(\gamma_3^2 k_1 - \gamma_4^2) - (\gamma_4^2 l_1 - \gamma_5^2)]). \quad (23f)$$

$$v_{3,1} = \frac{1}{3!}i_4(i_3[i_2(\gamma_1^3 i_1 - \gamma_2^3) - (\gamma_2^3 j_1 - \gamma_3^3)] - [j_2(\gamma_2^3 j_1 - \gamma_3^3) - (\gamma_3^3 k_1 - \gamma_4^3)]) - \frac{1}{3!}(j_3[j_2(\gamma_2^3 j_1 - \gamma_3^3) - (\gamma_3^3 k_1 - \gamma_4^3)] - [k_2(\gamma_3^3 k_1 - \gamma_4^3) - (\gamma_4^3 l_1 - \gamma_5^3)]). \quad (23g)$$

$$i_2 = \frac{b_6 j - c_6}{a_6 i_1 - b_6}, \qquad j_2 = \frac{c_6 k_1 - d_6}{b_6 j_1 - c_6}, \qquad k_2 = \frac{d_6 l_1 - e_6}{c_6 k_1 - d_6}, \quad (23h)$$

$$i_3 = \frac{j_2(b_5 j_1 - c_5) - (c_5 k_1 - d_5)}{i_2(a_5 i_1 - b_5) - (b_5 j_1 - c_5)}, \qquad j_3 = \frac{k_2(c_5 k_1 - d_5) - (d_5 l_1 - e_5)}{j_2(b_5 j_1 - c_5) - (c_5 k_1 - d_5)}, \quad (23i)$$

$$i_4 = \frac{j_3[j_2(b_4 j_1 - c_4) - (c_4 k_1 - d_4)] - [k_2(c_4 k_1 - d_4) - (d_4 l_1 - e_4)]}{i_3[i_2(a_4 i_1 - b_4) - (b_4 j_1 - c_4)] - [j_2(b_4 j_1 - c_4) - (c_4 k_1 - d_4)]}. \quad (23j)$$

Here the truncation error is given as

$$R = C_{\gamma_i} h^8 + O(h^9), \tag{23k}$$

where

$$C_{\gamma_i} = \frac{1}{8!} i_4 (i_3 [i_2 (\gamma_1^8 i_1 - \gamma_2^8) - (\gamma_2^8 j_1 - \gamma_3^8)] - [j_2 (\gamma_2^8 j_1 - \gamma_3^8) - (\gamma_3^8 k_1 - \gamma_4^8)]) \\ - \frac{1}{8!} (j_3 [j_2 (\gamma_2^8 j_1 - \gamma_3^8) - (\gamma_3^8 k_1 - \gamma_4^8)] - [k_2 (\gamma_3^8 k_1 - \gamma_4^8) - (\gamma_4^8 l_1 - \gamma_5^8)]). \tag{23l}$$

Notice that the parameter on the right-hand side is independent of $h$. It is important to further mention that even though we have $O(h^8)$ convergence, it is up to a constant factor $C_{\gamma_i}$ which is $\gamma_i$ dependent. For a uniform scheme, this constant is fixed. However, for the present high-order staggered scheme and observing ($21b$) and ($23h$)-($23j$), it can easily be seen $C_{\gamma_i}$ can vary substantially based on the controlling and adjustable factors $\gamma_1$, $\gamma_2$, $\gamma_3, \gamma_4$, and $\gamma_5$ and the latter represents the grid points distribution. The constant factor $C_{\gamma_i}$, depending on how small or large it is, can enable us to obtain a more or less accurate numerical solution with coarse grids. However, asymptotically, the step size still controls the narrative in terms of convergence rate. Finding an optimal grid points distribution that will minimize $C_{\gamma_i}$ combined with a high-order staggered boundary scheme can substantially enable a much more accurate numerical solution with a large step size. This process could further be trained to be adaptive. We hope to investigate this phenomenon in our future work. Heuristically, a simple computation reveals the following parameters for $C_{\gamma_i}$ based on the grid points distribution

$$C_{\gamma_i}(2,3,4,5,6) = 0.77143, \qquad C_{\gamma_i}(2,4,5,6,7) = 1.78646, \qquad C_{\gamma_i}(2,4,6,8,10) = 95.23810. \tag{24}$$

From (24), it can easily be seen that the value of $C_{\gamma_i}$ for grid points distribution $(2,3,4,5,6)$ is the smallest. This gain can improve accuracy in a non-asymptotic scenario when the grid is coarse. We provide our further observation based on this positive indication in the numerical example.

Furthermore, we also observed that shifting away from the left boundary at least up to the first two grid points when implementing Taylor series expansion for computing the optimal exercise boundary provides a more efficient result. Hence, we always select $\bar{x} \geq 2h$ and $\bar{x} \ll x$. If we choose a uniform step size, we will obtain the following uniform fifth-order boundary scheme as follows:

$$625(\bar{x},\cdot) - \frac{625}{4} f(2\bar{x},\cdot) + \frac{1024}{27} f(3\bar{x},\cdot) - \frac{625}{64} f(4\bar{x},\cdot) + f(5\bar{x},\cdot) \\ = \frac{874853}{1728\bar{x}} f(0,\cdot) + \frac{60095\bar{x}}{144} f'(0,\cdot) + \frac{3425\bar{x}^2}{24} f''(0,\cdot) + \frac{125\bar{x}^3}{6} f'''(0,\cdot) + O(h^8). \tag{25}$$

The motivation for implementing the present high order dynamic and staggered scheme is to enable us to select grid points very close to the left boundary in such a way that we still shift away from the left boundary at least up to the first two grid points when implementing Taylor series expansion.

To compute the optimal exercise boundary with precise accuracy using our high order staggered boundary scheme, we consider the square root transformation first presented in the work of Kim et al. [15] for solving American options with front-fixing. We further refer the reader to the work of Kim et al. [15, 16], Lee [17, 18], and Nwankwo and Dai [23-25] on various implementations of this transformation for improving accuracy and recovering convergence rate when pricing American style options. The square root function with fixed free-boundary is presented as follows:

$$Q(x>0,t)=\sqrt{U(x,t)-E+e^{x}s_f},\qquad Q(x\le 0,t)=0. \tag{26}$$

Kim et al [15] and Lee [17] analyzed the characteristic of $Q(x,\tau)$ and show that the function has a Lipschitz character and exhibits non-degeneracy and non-singularity near the optimal exercise boundary. It is well known that these degeneracy and singularity deteriorate the accuracy of American options. Here, we consider up to the third derivative of this function at the left boundary point with

$$Q(0,t)=0,\qquad Q'(0,t)=\frac{\sqrt{rE}}{\sigma},\qquad Q''(0,t)=-\frac{2\beta_t\sqrt{rE}}{3\sigma^3},\qquad Q'''(0,t)=\frac{2\beta_t^2\sqrt{rE}}{3\sigma^5}+\frac{r\sqrt{rE}}{2\sigma^3}. \tag{27}$$

Substituting (27) into the right-hand side of ($23a$), we then obtain as follows:

$$\frac{s_{1,1}}{h^2}f(h_1,\cdot)-\frac{s_{1,2}}{h^2}f(h_2,\cdot)+\frac{s_{1,3}}{h^2}f(h_3,\cdot)-\frac{s_{1,4}}{h^2}f(h_4,\cdot)+\frac{s_{1,5}}{h^2}f(h_5,\cdot)$$
$$=v_{1,1}\frac{\sqrt{rE}}{\sigma h}-v_{2,1}\frac{2\beta_t\sqrt{rE}}{3\sigma^3}+hv_{3,1}\left(\frac{2\beta_t^2\sqrt{rE}}{3\sigma^5}+\frac{r\sqrt{rE}}{2\sigma^3}\right)+O(h^5). \tag{28}$$

The optimal exercise boundary is then computed as follows:

$$\frac{ds_f}{dt}=g_{h,t}s_f,\qquad g_{h,t}=\frac{-\varpi-\sqrt{\varpi^2-4\alpha\kappa_{h,t}}}{2\alpha}; \tag{29a}$$

$$\alpha=\frac{h^3\sqrt{rE}}{3\sigma^5 s_f^2}v_{3,1},\qquad \varpi=-\frac{2h^2\sqrt{rE}}{3\sigma^3 s_f}v_{2,1}+\frac{4h^3\sqrt{rE}}{3\sigma^5 s_f}v_{3,1}; \tag{29b}$$

$$\kappa_{h,t}=-M_{5,1}+\frac{h\sqrt{rE}}{\sigma}v_{1,1}-\frac{h^2\left(r-\frac{\sigma^2}{2}\right)\sqrt{rE}}{3\sigma^3}v_{2,1}+\frac{2h^3\left(r-\frac{\sigma^2}{2}\right)^2\sqrt{rE}}{3\sigma^5}v_{3,1}+\frac{h^3r\sqrt{rE}}{2\sigma^3}v_{3,1}, \tag{29c}$$

$$M_{5,1}=s_{1,1}f(h_1,\cdot)-s_{2,1}f(h_2,\cdot)+s_{3,1}f(h_3,\cdot)-s_{4,1}f(h_4,\cdot)+s_{5,1}f(h_5,\cdot). \tag{29d}$$

Furthermore, we can generate another high-order staggered boundary scheme with fewer stencils. To this end, following a similar approach as presented in (20)-(24), we obtain as follows:

$$-\frac{s_{0,2}}{h^2}f(0,\cdot)+\frac{s_{1,2}}{h^2}f(h_1,\cdot)-\frac{s_{2,2}}{h^2}f(h_2,\cdot)+\frac{s_{3,2}}{h^2}f(h_3,\cdot)-\frac{s_{4,2}}{h^2}f(h_4,\cdot)+\frac{s_{5,2}}{h^2}f(h_5,\cdot)$$
$$=\frac{v_{1,2}f'(0,\cdot)}{h}+v_{2,2}f''(0,\cdot)+O(h^5), \qquad (30a)$$

$$i_5=\frac{b_6}{a_6}, \qquad j_5=\frac{c_6}{b_6}, \qquad k_5=\frac{d_6}{c_6}, \qquad l_5=\frac{e_6}{d_6}. \qquad (30b)$$

$$i_6=\frac{b_5j_5-c_5}{a_5i_5-b_5}, \qquad j_6=\frac{c_5k_5-d_5}{b_5j_5-c_5}, \qquad k_6=\frac{d_5l_5-e_5}{c_5k_5-d_5}, \qquad (30c)$$

$$i_7=\frac{j_6(b_4j_5-c_4)-(c_4k_5-d_4)}{i_6(a_4i_5-b_4)-(b_4j_5-c_4)}, \qquad j_7=\frac{k_6(c_4k_5-d_4)-(d_4l_5-e_4)}{j_6(b_4j_5-c_4)-(c_4k_5-d_4)}, \qquad (30d)$$

$$i_8=\frac{j_7[j_6(b_3j_5-c_3)-(c_3k_5-d_3)]-[k_6(c_3k_5-d_3)-(d_3l_5-e_3)]}{i_7[i_6(a_3i_5-b_3)-(b_3j_5-c_3)]-[j_6(b_3j_5-c_3)-(c_3k_5-d_3)]}. \qquad (30e)$$

$$s_{0,2}=i_8(i_7[i_6(\gamma_1i_5-\gamma_2)-(\gamma_2j_5-\gamma_3)]-[j_6(\gamma_2j_5-\gamma_3)-(\gamma_3k_5-\gamma_4)])$$
$$-(j_7[j_6(\gamma_2j_5-\gamma_3)-(\gamma_3k_5-\gamma_4)]-[k_6(\gamma_3k_5-\gamma_4)-(\gamma_4l_5-\gamma_5)]), \qquad (30f)$$

$$s_{1,2}=i_8i_7i_6i_5, \qquad s_{2,2}=\ i_8i_7i_6+i_8i_7j_5+i_7j_6j_5+j_7j_6j_5; \qquad (30g)$$

$$s_{3,2}=i_8i_7+i_8j_6+i_8k_5+j_7k_5+j_7j_6+k_6k_5, \qquad s_{4,2}=i_8+j_7+k_6+l_5, \qquad s_{5,2}=1; \qquad (30h)$$

$$v_{1,2}=i_8(i_7[i_6(\gamma_1i_5-\gamma_2)-(\gamma_2j_5-\gamma_3)]-[j_6(\gamma_2j_5-\gamma_3)-(\gamma_3k_5-\gamma_4)])$$
$$-(j_7[j_6(\gamma_2j_5-\gamma_3)-(\gamma_3k_5-\gamma_4)]-[k_6(\gamma_3k_5-\gamma_4)-(\gamma_4l_5-\gamma_5)]). \qquad (30i)$$

$$v_{2,2}=\frac{1}{2!}i_8(i_7[i_6(\gamma_1^2i_5-\gamma_2^2)-(\gamma_2^2j_5-\gamma_3^2)]-[j_6(\gamma_2^2j_5-\gamma_3^2)-(\gamma_3^2k_5-\gamma_4^2)])$$
$$-\frac{1}{2!}(j_7[j_6(\gamma_2^2j_5-\gamma_3^2)-(\gamma_3^2k_5-\gamma_4^2)]-[k_6(\gamma_3^2k_5-\gamma_4^2)-(\gamma_4^2l_5-\gamma_5^2)]). \qquad (30j)$$

Subtracting $(24a)$ from $(30a)$, we then obtain the second staggered high-order boundary scheme

$$-\frac{(s_{0,1}-s_{0,2})}{h^2}f(0,\cdot)+\frac{(s_{1,1}-s_{1,2})}{h^2}f(h_1,\cdot)-\frac{(s_{2,1}-s_{2,2})}{h^2}f(h_2,\cdot)+\frac{(s_{3,1}-s_{3,2})}{h^2}f(h_3,\cdot)$$
$$-\frac{(s_{4,1}-s_{4,2})}{h^2}f(h_4,\cdot)=\frac{(v_{1,1}-v_{1,2})}{h}f'(0,\cdot)+(v_{2,1}-v_{2,2})f''(0,\cdot)+v_{3,1}hf'''(0,\cdot)$$
$$+O(h^5). \qquad (30k)$$

Substituting (27) into the right-hand side of $(30k)$

$$\frac{(s_{1,1}-s_{1,2})}{h^2}f(h_1,\cdot)-\frac{(s_{2,1}-s_{2,2})}{h^2}f(h_2,\cdot)+\frac{(s_{3,1}-s_{3,2})}{h^2}f(h_3,\cdot)-\frac{(s_{4,1}-s_{4,2})}{h^2}f(h_4,\cdot)$$

$$=(v_{1,1}-v_{1,2})\frac{\sqrt{rE}}{\sigma h}-(v_{2,1}-v_{2,2})\frac{2\beta_t\sqrt{rE}}{3\sigma^3}+hv_{3,1}\left(\frac{2\beta_t^2\sqrt{rE}}{3\sigma^5}+\frac{r\sqrt{rE}}{2\sigma^3}\right)+O(h^5). \qquad (30l)$$

Like (29), the optimal exercise can be computed from (30$l$) with high accuracy.

**Remark 1**. It is further important to mention that the present high-order staggered boundary schemes with 4 and 5 stencils can be used to generate arbitrary weights based on the selected grid points distribution. It gives us control of the non-uniformity of our scheme and the choice of grid points by simply adjusting the parameter $\gamma_1$, $\gamma_2, \gamma_3, \gamma_4$, and $\gamma_5$. This approach can further be extended to generate schemes with lower-order accuracy when a low-order interior scheme is implemented.

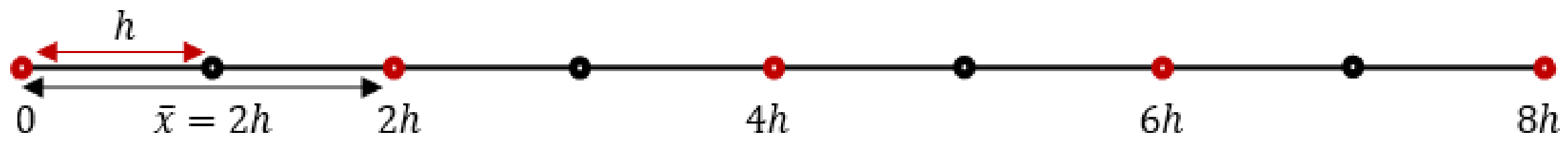

**Fig. 1a** High order boundary scheme with uniform grid points $\bar{x} = 2h$. The grid points in red represent those involved in the computation of the optimal exercise boundary.

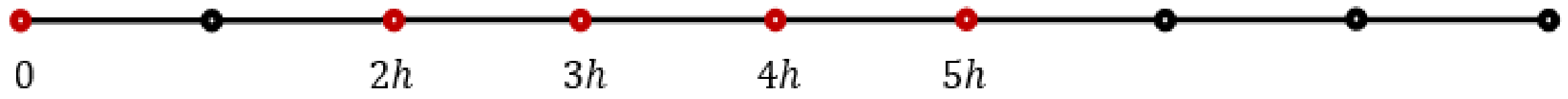

**Fig. 1b.** High order staggered boundary scheme with grid points distribution $(\gamma_1, \gamma_2, \gamma_3, \gamma_4) = (2,3,4,5)$. The grid points in red represent those involved in the computation of the optimal exercise boundary.

### 2.3. 3(2) RK-Bogacki and Shampine Adaptive Time-Stepping

Several Runge-Kutta embedded pairs have been propose and implemented in works of literature [3-5, 9-11, 14, 19, 23-26, 28-30, 32, 35]. Here, we consider 3(2) Bogacki-Shampine embedded pairs [2] which are third-order accurate. The authors argue that it outperforms other existing third-order accurate methods with the second-order embedded method. Furthermore, efficient implementation of 5(4) embedded pairs for solving systems of diffusion-convective-reactive partial differential equations arising in the American options problem was described in the works of Nwankwo and Dai [23-25]. We further refer the reader to their works. Here, we slightly present some minor enhancements. The computational procedure for the implementation of the sixth-order compact scheme with 3(2) Bogacki-Shampine pairs is as follows:

With slight abuse of notation, let the two semi-discrete coupled systems of equations representing the asset option and delta sensitivity with the optimal exercise boundary be given as

$$\frac{\partial \boldsymbol{u}^n}{\partial t} = \boldsymbol{\mathcal{L}}_u^n, \qquad \frac{\partial \boldsymbol{w}^n}{\partial t} = \boldsymbol{\mathcal{L}}_w^n, \qquad \frac{ds_f^n}{dt} = \mathcal{L}_{s_f}^n; \tag{31a}$$

$$\boldsymbol{\mathcal{L}}_u^n = \frac{\sigma^2}{2} B_{5,h}^{-\mathbf{1}}(A\boldsymbol{u}^n + \boldsymbol{f}_u^n) + \beta^n(\tau)\boldsymbol{w}^n - r\boldsymbol{u}^n, \tag{31b}$$

$$\boldsymbol{\mathcal{L}}_w^n = \frac{\sigma^2}{2} B_{5,h}^{-\mathbf{1}}(A\boldsymbol{w}^n + \boldsymbol{f}_w^n) + \beta^n(\tau) B_{5,h}^{-\mathbf{1}}(A\boldsymbol{u}^n + \boldsymbol{f}_w^n) - r\boldsymbol{w}^n. \tag{31c}$$

1st Stage:

$$\mathcal{L}_{s_f}^n = g_h^n s_f^n, \qquad \beta^n = \frac{\mathcal{L}_{s_f}^n}{s_f^n} + \nu, \qquad \nu = r - \frac{1}{2}\sigma^2;$$

$$\boldsymbol{\mathcal{L}}_u^n = \frac{\sigma^2}{2} B_{5,h}^{-\mathbf{1}}(A\boldsymbol{u}^n + \boldsymbol{f}_u^n) + \beta_t^n \boldsymbol{w}^n - r\boldsymbol{u}^n,$$

$$\boldsymbol{\mathcal{L}}_w^n = \frac{\sigma^2}{2} B_{5,h}^{-\mathbf{1}}(A\boldsymbol{w}^n + \boldsymbol{f}_w^n) + \beta^n B_{5,h}^{-\mathbf{1}}(A\boldsymbol{u}^n + \boldsymbol{f}_w^n) - r\boldsymbol{w}^n,$$

$$s_f^{n+\frac{1}{4}} = s_f^n + \frac{1}{2}k\mathcal{L}_{s_f}^n, \qquad \boldsymbol{u}^{n+\frac{1}{4}} = \boldsymbol{u}^n + \frac{1}{2}k\boldsymbol{\mathcal{L}}_u^n, \qquad \boldsymbol{w}^{n+\frac{1}{4}} = \boldsymbol{w}^n + \frac{1}{2}k\boldsymbol{\mathcal{L}}_w^n.$$

2nd Stage:

$$\mathcal{L}_{s_f}^{n+\frac{1}{4}} = g_h^{n+\frac{1}{4}} s_f^{n+\frac{1}{4}}, \qquad \beta^{n+\frac{1}{4}} = \frac{\mathcal{L}_{s_f}^{n+\frac{1}{4}}}{s_f^{n+\frac{1}{4}}} + \nu;$$

$$\boldsymbol{\mathcal{L}}_u^{n+\frac{1}{4}} = \frac{\sigma^2}{2} B_{5,h}^{-\mathbf{1}}\left(A\boldsymbol{u}^{n+\frac{1}{4}} + \boldsymbol{f}_u^{n+\frac{1}{4}}\right) + \beta_t^{n+\frac{1}{4}}\boldsymbol{w}^{n+\frac{1}{4}} - r\boldsymbol{u}^{n+\frac{1}{4}},$$

$$\boldsymbol{\mathcal{L}}_w^{n+\frac{1}{4}} = \frac{\sigma^2}{2} B_{5,h}^{-\mathbf{1}}\left(A\boldsymbol{w}^{n+\frac{1}{4}} + \boldsymbol{f}_w^{n+\frac{1}{4}}\right) + \beta^{n+\frac{1}{4}} B_{5,h}^{-\mathbf{1}}\left(A\boldsymbol{u}^{n+\frac{1}{4}} + \boldsymbol{f}_w^{n+\frac{1}{4}}\right) - r\boldsymbol{w}^{n+\frac{1}{4}},$$

$$s_f^{n+\frac{2}{4}} = s_f^n + \frac{3}{4}k\mathcal{L}_{s_f}^{n+\frac{1}{4}}, \qquad \boldsymbol{u}^{n+\frac{2}{4}} = \boldsymbol{u}^n + \frac{3}{4}k\boldsymbol{\mathcal{L}}_u^{n+\frac{1}{4}}, \qquad \boldsymbol{w}^{n+\frac{2}{4}} = \boldsymbol{w}^n + \frac{3}{4}k\boldsymbol{\mathcal{L}}_w^{n+\frac{1}{4}};$$

3rd Stage:

$$\mathcal{L}_{s_f}^{n+\frac{2}{4}} = g_h^{n+\frac{2}{4}} s_f^{n+\frac{2}{4}}, \qquad \beta^{n+\frac{2}{4}} = \frac{\mathcal{L}_{s_f}^{n+\frac{2}{4}}}{s_f^{n+\frac{2}{4}}} + \nu;$$

$$\boldsymbol{\mathcal{L}}_u^{n+\frac{2}{4}} = \frac{\sigma^2}{2} B_{5,h}^{-\mathbf{1}}\left(A\boldsymbol{u}^{n+\frac{2}{4}} + \boldsymbol{f}_u^{n+\frac{2}{4}}\right) + \beta_t^{n+\frac{2}{4}}\boldsymbol{w}^{n+\frac{2}{4}} - r\boldsymbol{u}^{n+\frac{2}{4}},$$

$$\boldsymbol{\mathcal{L}}_w^{n+\frac{2}{4}} = \frac{\sigma^2}{2} B_{5,h}^{-\mathbf{1}}\left(A\boldsymbol{w}^{n+\frac{2}{4}} + \boldsymbol{f}_w^{n+\frac{2}{4}}\right) + \beta^{n+\frac{2}{4}} B_{5,h}^{-\mathbf{1}}\left(A\boldsymbol{u}^{n+\frac{2}{4}} + \boldsymbol{f}_w^{n+\frac{2}{4}}\right) - r\boldsymbol{w}^{n+\frac{2}{4}},$$

$$s_f^{n+\frac{3}{4}} = s_f^n + \frac{2}{9}k\mathcal{L}_{s_f}^n + \frac{1}{3}k\mathcal{L}_{s_f}^{n+\frac{1}{4}} + \frac{4}{9}k\mathcal{L}_{s_f}^{n+\frac{2}{4}},$$

$$\boldsymbol{u}^{n+\frac{3}{4}} = \boldsymbol{u}^n + \frac{2}{9}k\boldsymbol{\mathcal{L}}_u^n + \frac{1}{3}k\boldsymbol{\mathcal{L}}_u^{n+\frac{1}{4}} + \frac{4}{9}k\boldsymbol{\mathcal{L}}_u^{n+\frac{2}{4}}, \qquad \boldsymbol{w}^{n+\frac{3}{4}} = \boldsymbol{w}^n + \frac{2}{9}k\boldsymbol{\mathcal{L}}_w^n + \frac{1}{3}k\boldsymbol{\mathcal{L}}_w^{n+\frac{1}{4}} + \frac{4}{9}k\boldsymbol{\mathcal{L}}_w^{n+\frac{2}{4}}.$$

Last Stage:

$$\mathcal{L}_{s_f}^{n+\frac{3}{4}} = g_h^{n+\frac{3}{4}} s_f^{n+\frac{3}{4}}, \qquad \beta^{n+\frac{3}{4}} = \frac{\mathcal{L}_{s_f}^{n+\frac{3}{4}}}{s_f^{n+\frac{3}{4}}} + \nu;$$

$$\boldsymbol{\mathcal{L}}_u^{n+\frac{3}{4}} = \frac{\sigma^2}{2} B_{5,h}^{-1}\left(A\boldsymbol{u}^{n+\frac{3}{4}} + \boldsymbol{f}_u^{n+\frac{3}{4}}\right) + \beta_t^{n+\frac{3}{4}}\boldsymbol{w}^{n+\frac{3}{4}} - r\boldsymbol{u}^{n+\frac{3}{4}},$$

$$\boldsymbol{\mathcal{L}}_w^{n+\frac{3}{4}} = \frac{\sigma^2}{2} B_{5,h}^{-\boldsymbol{1}}\left(A\boldsymbol{w}^{n+\frac{3}{4}} + \boldsymbol{f}_w^{n+\frac{3}{4}}\right) + \beta^{n+\frac{3}{4}} B_{5,h}^{-\boldsymbol{1}}\left(A\boldsymbol{u}^{n+\frac{3}{4}} + \boldsymbol{f}_w^{n+\frac{3}{4}}\right) - r\boldsymbol{w}^{n+\frac{3}{4}},$$

$$s_f^{n+1} = s_f^{n+\frac{3}{4}}, \qquad \bar{s}_f^{n+1} = s_f^n + \frac{k}{24}\left(7\mathcal{L}_{s_f}^n + 6\mathcal{L}_{s_f}^{n+\frac{1}{4}} + 8\mathcal{L}_{s_f}^{n+\frac{2}{4}} + 3\mathcal{L}_{s_f}^{n+\frac{3}{4}}\right),$$

$$\boldsymbol{u}^{n+1} = \boldsymbol{u}^{n+\frac{3}{4}}, \qquad \bar{\boldsymbol{u}}^{n+1} = \boldsymbol{u}^n + \frac{k}{24}\left(7\boldsymbol{\mathcal{L}}_u^n + 6\boldsymbol{\mathcal{L}}_u^{n+\frac{1}{4}} + 8\boldsymbol{\mathcal{L}}_u^{n+\frac{2}{4}} + 3\boldsymbol{\mathcal{L}}_u^{n+\frac{3}{4}}\right),$$

$$\boldsymbol{w}^{n+1} = \boldsymbol{w}^{n+\frac{3}{4}}, \qquad \bar{\boldsymbol{w}}^{n+1} = \boldsymbol{w}^n + \frac{k}{24}\left(7\boldsymbol{\mathcal{L}}_w^n + 6\boldsymbol{\mathcal{L}}_w^{n+\frac{1}{4}} + 8\boldsymbol{\mathcal{L}}_w^{n+\frac{2}{4}} + 3\boldsymbol{\mathcal{L}}_w^{n+\frac{3}{4}}\right).$$

An error threshold is defined as follows:

$$e_u = \|\bar{\boldsymbol{u}}^{n+1} - \boldsymbol{u}^{n+1}\|_\infty, \qquad e_w = \|\bar{\boldsymbol{w}}^{n+1} - \boldsymbol{w}^{n+1}\|_\infty, \qquad e_{s_f} = \left|\bar{s}_f^{n+1} - s_f^{n+1}\right|, \tag{32a}$$

such that we update the new time step at each time level based on the criterion below

$$k_{new} = \begin{cases} \varrho k_{old}\left[\varepsilon/\max\left(e_u, e_w, e_{s_f}\right)\right]^{1/2}, & \max\left(e_u, e_w, e_{s_f}\right) < \varepsilon, \\ \varrho k_{old}\left[\varepsilon/\max\left(e_u, e_w, e_{s_f}\right)\right]^{1/3}, & \max\left(e_u, e_w, e_{s_f}\right) \geq \varepsilon. \end{cases} \tag{32b}$$

Here, $\varepsilon$ is a given tolerance. $k_{new}$ is optimal if $\max\left(e_u, e_w, e_{s_f}\right) < \varepsilon$ and as such, $k_{new} = k$ and the obtained numerical solutions of the optimal exercise boundary, asset option, and delta sensitivity are accepted. If $\max\left(e_u, e_w, e_{s_f}\right) \geq \varepsilon$, we obtain a new time step based on (32) and use the latter to obtain a new solution. The process will continue till $\max\left(e_u, e_w, e_{s_f}\right) < \varepsilon$ is achieved. Furthermore, in the numerical example section, we varied $\varrho$ and observe its effect on the computational time and accuracy of our numerical approximations.

**Derivatives of the Optimal Exercise Boundary**

In this research, we consider the numerical solutions of the optimal exercise boundary and its derivatives. In essence, we are more interested to investigate the profile of the derivatives of the optimal exercise after improving it using the high-order analytical approximation as presented in subsection 2.2. We consider up to the second derivative of the optimal exercise boundary. The numerical solution of the first derivative of the optimal exercise boundary is easily obtained from (29) or (30). For the second derivative of the optimal exercise boundary, we need to consider some extra derivations as follows:

$$U(x,t) = Q^2(x,t) + E - e^x s_f(t), \tag{33a}$$

$$U_{xxx}(0,t) = -\frac{4\left(r - \frac{1}{2}\sigma^2\right)rE}{\sigma^4} - \frac{4rE}{\sigma^4 s_f(t)}\frac{ds_f(t)}{dt} - s_f(t), \tag{33b}$$

$$U_{xxxt}(0,t) = -\frac{4rE}{\sigma^4 s_f(t)}\frac{d^2 s_f(t)}{dt^2} + \frac{4rE}{\sigma^4 s_f^2(t)}\left(\frac{ds_f(t)}{dt}\right)^2 - \frac{ds_f(t)}{dt}, \tag{33c}$$

$$U_{xxxx}(0,t) = \frac{8\beta_t^2 rE}{\sigma^6} + \frac{4r^2E}{\sigma^4} - s_f, \tag{33d}$$

$$U_{xxxxx}(0,t) = -\frac{80\beta_t^3 rE}{9\sigma^8} - \frac{20\beta_t r^2 E}{3\sigma^6} + \frac{10\sqrt{rE}}{\sigma}Q_{xxxx}(0,t) - s_f. \tag{33e}$$

Differentiating the transformed American option model in (3) thrice and substituting (33b)−(33e), we then obtain

$$-\frac{4rE}{\sigma^4 s_f(t)}\frac{d^2 s_f(t)}{dt^2} + \frac{4rE}{\sigma^4 s_f^2(t)}\left(\frac{ds_f(t)}{dt}\right)^2 - \frac{ds_f(t)}{dt} - \frac{\sigma^2}{2}\left(-\frac{80\beta_t^3 rE}{9\sigma^8} - \frac{20\beta_t r^2E}{3\sigma^6} + \frac{10\sqrt{rE}}{\sigma}Q_{xxxx}(0,t) - s_f\right) + \beta_t\left(\frac{8\beta_t^2 rE}{\sigma^6} + \frac{4r^2E}{\sigma^4} - s_f\right) + r\left(-\frac{4rE}{\sigma^4}\beta_t - s_f(t)\right). \tag{33f}$$

Solving for $Q_{xxxx}(0,t)$, we then obtain

$$Q_{xxxx}(0,t) = -\frac{\sqrt{rE}}{\sigma^5 s_f(t)}\frac{d^2 s_f(t)}{dt^2} + \frac{\sqrt{rE}}{\sigma^5 s_f^2(t)}\left(\frac{ds_f(t)}{dt}\right)^2 - \frac{32\beta_t^3 rE}{45\sigma^7} - \frac{14\beta_t r^2 E}{15\sigma^5}. \tag{33g}$$

Furthermore, as mentioned earlier, $\bar{x} \ll x$ and if we chose $\bar{x} = 2h$, another high-order analytical approximation for approximating the second derivative of the optimal exercise boundary is introduced as follows:

$$1024Q(\bar{x},\cdot) - 96Q(2\bar{x},\cdot) + \frac{1024}{81}Q(3\bar{x},\cdot) - Q(4\bar{x},\cdot) = \frac{23380\bar{x}}{27}Q_x(0,\cdot) + \frac{3320\bar{x}^2}{9}Q_{xx}(0,\cdot) + \frac{800\bar{x}^3}{9}Q_{xxx}(0,\cdot) + \frac{32\bar{x}^4}{3}Q_{xxxx}(0,\cdot) + O(h^8). \quad (33h)$$

We then compute the second derivative of the optimal exercise boundary as follows:

$$\frac{d^2 s_f(t)}{dt^2} = -\rho\left[M_{4,h} - \frac{23380\bar{x}\sqrt{rE}}{27\sigma} + \frac{6640\bar{x}^2\sqrt{rE}}{27\sigma^3}\beta_t - \frac{800\bar{x}^3}{9}\left(\frac{2\beta_t^2\sqrt{rE}}{3\sigma^5} + \frac{r\sqrt{rE}}{2\sigma^3}\right) - \frac{32\bar{x}^4}{3}\left(\frac{\sqrt{rE}}{\sigma^5 s_f^2(t)}\left(\frac{ds_f(t)}{dt}\right)^2 - \frac{32\beta_t^3 rE}{45\sigma^7} - \frac{14\beta_t r^2 E}{15\sigma^5}\right)\right], \qquad \rho = \frac{\sigma^5 s_f(t)}{\sqrt{rE}}. \quad (33i)$$

Because the equations for computing the first and second-order derivative of the optimal exercise boundary do not directly involve time discretization, we can easily compute the first- and second-order derivative of the optimal exercise boundary using the numerical solutions of the asset option at $n$ −time level based on (29) or (30) and (33).

## 3. Numerical Examples

In this section, we consider a couple of numerical examples to verify and validate the performance of our present method with respect to computational cost, numerical accuracy, and recovering numerical convergence rate that is in good agreement with the theoretical convergence rate. To this end, we consider the following examples with short, medium, and a long time to maturity.

$$E = 100, \qquad r = 0.05, \qquad \sigma = 0.2, \qquad T = 0.5. \quad (34a)$$

$$E = 100, \qquad r = 0.1, \qquad \sigma = 0.3, \qquad T = 1. \quad (34b)$$

$$E = 100, \qquad r = 0.08, \qquad \sigma = 0.2, \qquad T = 3. \quad (34c)$$

Here, we use varying large tolerances and step sizes and adapt our time stepping to be optimal at each time level based on the given tolerance and step size. Our code is written in MATLAB R2022b with a laptop speed of 1.7GHz. For convenience, we label our method as follows:

- Sixth-order compact scheme with fifth-order near-boundary scheme and 5-stencil fifth-order staggered boundary scheme – $CS - 55\ (\gamma_1, \gamma_2, \gamma_3, \gamma_4, \gamma_5)$.
- Sixth-order compact scheme with fifth-order near-boundary scheme and 4-stencil fifth-order staggered boundary scheme – $CS - 54\ (\gamma_1, \gamma_2, \gamma_3, \gamma_4)$.

First, we computed the convergence rate of our numerical scheme using the examples ($34a$) and verify how well it agrees with its theoretical counterpart. Because our adaptive scheme is third-order accurate, we computed the convergence rate of our numerical scheme using SSPRK3 which has strong stability property [12, 27] and is third-order accurate. We computed the convergence rate of the asset option, delta sensitivity, optimal exercise boundary, and the derivative of the optimal exercise boundary with $k = 10^{-6}$. The results were displayed in Table 2.

From Table 2, one can easily see that the obtained convergence rate is in very good agreement with the implemented scheme as the step size decreases. It is better than the one we obtained in our previous work [23-25] where we implemented a fourth-order compact scheme. Also, a very high convergence rate was obtained for the first derivative of the optimal exercise boundary. This is because we computed the latter with the high-order analytical approximation in ($20a$) or ($30l$). Furthermore, in most cases, the error decreases a bit faster in grid points distributions $(\gamma_1, \gamma_2, \gamma_1, \gamma_2) = (2,3,4,5)$ and $(2,3,4,5,6)$ when compared with $(\gamma_1, \gamma_2, \gamma_1, \gamma_2) = (2,4,5,6), (2,4,5,6,7), (2,4,6,8)$ and $(2,4,6,8,10)$.

Next, we computed the numerical solution of the asset option using the example in ($34c$). The benchmark value was obtained from the work of Cox et al. [8]. The results are displayed in Table 3. We also computed the numerical solutions of the optimal exercise boundary and its derivatives with varying distributions of grid points. The numerical solutions are listed in Table 4 using the example in ($34b$). We display the plot profile of the optimal exercise boundary with its first and second derivatives in Figure 2 based on the example in (34c).

We observe from Tables 3 and 4 that the obtained results are very close to the benchmark value with very coarse grids. Furthermore, when the grid is very coarse, we observe that the numerical solution from the grid points distribution $(\gamma_1, \gamma_2, \gamma_1, \gamma_2) = (3,4,5,6)$ is more accurate when compared with the solution obtained from $(\gamma_1, \gamma_2, \gamma_1, \gamma_2) = (3,6,9,12)$. Furthermore, In Table 4, it is important to observe how the numerical solution of the optimal exercise boundary and its derivatives behave based on the grid points distribution using the high-order staggered boundary scheme we presented in this work. For instance, the numerical solution using $(\gamma_1, \gamma_2, \gamma_1, \gamma_2) = (2,3,4,5)$ is almost the same as the one obtained from $(\gamma_1, \gamma_2, \gamma_1, \gamma_2) = (3,4,5,6)$. Also, the numerical solution obtained with grid points distribution $(\gamma_1, \gamma_2, \gamma_1, \gamma_2) = (2,4,6,8)$ is like the one obtained from $(\gamma_1, \gamma_2, \gamma_1, \gamma_2) = (3,5,7,9)$. As we have already mentioned, when the number of stencils increases, the high-order staggered boundary schemes leverage the position to keep away from the left boundary as much as possible and still use grid points very close to the latter for computing the boundary values and the optimal exercise boundary. Hence, enhancing the accuracy of the numerical approximation of the optimal exercise boundary, asset option, and their derivatives.

For instance, in our previous work [25] where we implemented a fourth-order compact finite difference scheme and fifth-order RK-Dormand and Prince embedded pairs (which is one of 5(4) Runge-Kutta pairs that provides more accurate result), we obtained $s_f(t) = 76.17$ when $h = 0.06$. However, as shown in Table 4 with grid points distribution $(\gamma_1, \gamma_2, \gamma_1, \gamma_2) = (2,3,4,5)$, we obtained exactly $s_f(t) = 76.16$ when $h = 0.06$. This shows that our sixth order scheme with the grid points distribution $(\gamma_1, \gamma_2, \gamma_1, \gamma_2) = (2,3,4,5)$ provides more accurate numerical solutions. In general, our numerical scheme yields highly accurate numerical approximations of the boundary values, asset option, and delta sensitivity with very coarse grids and large tolerance up to $\varepsilon = 10^{-2}$.

We also compared the total CPU time(s) and numerical accuracy between the SSPRK3 scheme and 3(2) Bogacki-Shampine pairs with varying large tolerances. Furthermore, we compared the total CPU time(s) and the accuracy of our numerical approximation with 3(2) Bogacki-Shampine pairs and varying $\varrho$. The results were listed in Tables 5a and 5b. For the computational time, when $h = 0.02$ and $\varrho = 0.03$ we observed from Table 5a that 3(2) Bogacki-Shampine embedded pairs is more than four times faster than SSPRK3 in achieving the same numerical accuracy. Furthermore, it implies that we can achieve greater accuracy with fast computation, very coarse grids, and large tolerance using 3(2) RK-Bogacki-Shampine pairs. It will be very useful in high dimensional context for improving computational cost.

Furthermore, in Table 5b with $h = 0.02$, it can be easily seen that when $\varrho$ is varied and for $\varrho = 0.3$, we obtained the smallest total CPU total time of $7.0$ seconds which is more than six times faster than SSPRK3 with the same accuracy. It is worth mentioning that we could further enhance our numerical result using higher-order embedded pairs. However, in this work, we focused on lower-order pairs which involve fewer functions evaluation.

**Table 2a.** Error and convergence rate in space with $CS-55$.

| | | (2,3,4,5,6) | | |
|---|---|---|---|---|
| $h$ | $maximum\ error$ | $convergence\ rate$ | $maximum\ error$ | $convergence\ rate$ |
| | $Asset\ option$ | | $Delta\ sensitivity$ | |
| 0.1 | ~ | ~ | ~ | ~ |
| 0.05 | $6.080\times10^{-1}$ | ~ | $4.683\times10^{0}$ | ~ |
| 0.025 | $4.807\times10^{-2}$ | 3.085 | $7.148\times10^{-1}$ | 2.712 |
| 0.0125 | $8.726\times10^{-4}$ | 6.359 | $1.177\times10^{-2}$ | 5.966 |
| 0.00625 | $2.643\times10^{-5}$ | 5.045 | $1.882\times10^{-4}$ | 5.924 |
| | $Optimal\ exercise\ boundary$ | | $First\ derivative\ of\ the\ optimal\ exercise\ boundary$ | |
| 0.1 | ~ | ~ | ~ | ~ |
| 0.05 | $6.080\times10^{-1}$ | ~ | $6.509\times10^{0}$ | ~ |
| 0.025 | $4.807\times10^{-2}$ | 3.661 | $9.320\times10^{-1}$ | 2.804 |
| 0.0125 | $2.879\times10^{-4}$ | 7.384 | $6.066\times10^{-3}$ | 7.263 |
| 0.00625 | $1.502\times10^{-5}$ | 4.261 | $2.487\times10^{-4}$ | 4.608 |
| | | (2,4,5,6,7) | | |
| | $Asset\ option$ | | $Delta\ sensitivity$ | |
| 0.1 | ~ | ~ | ~ | ~ |
| 0.05 | $1.322\times10^{0}$ | ~ | $7.833\times10^{1}$ | ~ |
| 0.025 | $6.621\times10^{-2}$ | 4.330 | $7.034\times10^{-1}$ | 3.477 |
| 0.0125 | $2.117\times10^{-3}$ | 4.967 | $1.623\times10^{-2}$ | 5.437 |
| 0.00625 | $9.130\times10^{-5}$ | 4.536 | $5.578\times10^{-4}$ | 4.863 |
| | $Optimal\ exercise\ boundary$ | | $First\ derivative\ of\ the\ optimal\ exercise\ boundary$ | |
| 0.1 | ~ | ~ | ~ | ~ |
| 0.05 | $1.919\times10^{0}$ | ~ | $1.047\times10^{1}$ | ~ |
| 0.025 | $4.854\times10^{-2}$ | 5.118 | $7.723\times10^{-1}$ | 3.761 |
| 0.0125 | $2.945\times10^{-3}$ | 4.456 | $5.129\times10^{-2}$ | 7.234 |
| 0.00625 | $1.327\times10^{-4}$ | 4.363 | $8.326\times10^{-5}$ | 5.945 |
| | | (2,4,6,8,10) | | |
| | $Asset\ option$ | | $Delta\ sensitivity$ | |
| 0.1 | ~ | ~ | ~ | ~ |
| 0.05 | $1.919\times10^{0}$ | ~ | $1.041\times10^{1}$ | ~ |
| 0.025 | $7.847\times10^{-2}$ | 4.612 | $8.004\times10^{-1}$ | 3.700 |
| 0.0125 | $3.241\times10^{-3}$ | 4.598 | $2.363\times10^{-2}$ | 5.081 |
| 0.00625 | $1.407\times10^{-4}$ | 4.525 | $8.436\times10^{-4}$ | 4.808 |
| | $Optimal\ exercise\ boundary$ | | $First\ derivative\ of\ the\ optimal\ exercise\ boundary$ | |
| 0.1 | ~ | ~ | ~ | ~ |
| 0.05 | $1.919\times10^{0}$ | ~ | $1.183\times10^{1}$ | ~ |
| 0.025 | $4.854\times10^{-2}$ | 5.304 | $6.956\times10^{-1}$ | 4.088 |
| 0.0125 | $2.945\times10^{-3}$ | 4.043 | $1.302\times10^{-2}$ | 5.739 |
| 0.00625 | $1.327\times10^{-4}$ | 4.472 | $2.758\times10^{-4}$ | 5.561 |

**Table 2b.** Error and convergence rate in space with $CS-54$.

| $h$ | $maximum\ error$ | $convergence\ rate$ | $maximum\ error$ | $convergence\ rate$ |
|---|---|---|---|---|
| | | (2,3,4,5) | | |
| | $Asset\ option$ | | $Delta\ sensitivity$ | |
| 0.1 | ~ | ~ | ~ | ~ |
| 0.05 | $1.111\times10^{0}$ | ~ | $6.890\times10^{1}$ | ~ |
| 0.025 | $7.607\times10^{-2}$ | 3.868 | $7.542\times10^{-1}$ | 3.192 |
| 0.0125 | $1.682\times10^{-3}$ | 5.499 | $1.330\times10^{-2}$ | 5.824 |
| 0.00625 | $6.193\times10^{-5}$ | 4.764 | $3.992\times10^{-4}$ | 5.059 |
| | $Optimal\ exercise\ boundary$ | | $First\ derivative\ of\ the\ optimal\ exercise\ boundary$ | |
| 0.1 | ~ | ~ | ~ | ~ |
| 0.05 | $1.111\times10^{0}$ | ~ | $9.188\times10^{0}$ | ~ |
| 0.025 | $5.047\times10^{-2}$ | 4.461 | $8.443\times10^{-1}$ | 3.444 |
| 0.0125 | $1.245\times10^{-3}$ | 5.341 | $9.840\times10^{-2}$ | 9.796 |
| 0.00625 | $5.416\times10^{-5}$ | 4.523 | $8.988\times10^{-4}$ | 3.448 |
| | | (2,4,5,6) | | |
| | $Asset\ option$ | | $Delta\ sensitivity$ | |
| 0.1 | ~ | ~ | ~ | ~ |
| 0.05 | $1.962\times10^{0}$ | ~ | $1.061\times10^{1}$ | ~ |
| 0.025 | $7.814\times10^{-2}$ | 4.650 | $8.087\times10^{-1}$ | 3.713 |
| 0.0125 | $3.567\times10^{-3}$ | 4.453 | $2.569\times10^{-2}$ | 4.976 |
| 0.00625 | $1.495\times10^{-4}$ | 4.577 | $8.988\times10^{-4}$ | 4.837 |
| | $Optimal\ exercise\ boundary$ | | $First\ derivative\ of\ the\ optimal\ exercise\ boundary$ | |
| 0.1 | ~ | ~ | ~ | ~ |
| 0.05 | $1.962\times10^{0}$ | ~ | $1.167\times10^{1}$ | ~ |
| 0.025 | $4.747\times10^{-2}$ | 5.369 | $8.213\times10^{-1}$ | 3.829 |
| 0.0125 | $3.291\times10^{-3}$ | 3.851 | $1.635\times10^{-2}$ | 5.650 |
| 0.00625 | $1.408\times10^{-4}$ | 4.547 | $2.938\times10^{-4}$ | 5.799 |
| | | (2,4,6,8) | | |
| | $Asset\ option$ | | $Delta\ sensitivity$ | |
| 0.1 | ~ | ~ | ~ | ~ |
| 0.05 | $2.389\times10^{0}$ | ~ | $1.245\times10^{1}$ | ~ |
| 0.025 | $8.134\times10^{-2}$ | 4.876 | $8.621\times10^{-1}$ | 3.852 |
| 0.0125 | $4.848\times10^{-3}$ | 4.068 | $3.368\times10^{-2}$ | 4.678 |
| 0.00625 | $2.020\times10^{-4}$ | 4.585 | $1.205\times10^{-3}$ | 4.085 |
| | $Optimal\ exercise\ boundary$ | | $First\ derivative\ of\ the\ optimal\ exercise\ boundary$ | |
| 0.1 | ~ | ~ | ~ | ~ |
| 0.05 | $2.389\times10^{0}$ | ~ | $1.192\times10^{1}$ | ~ |
| 0.025 | $4.758\times10^{-2}$ | 5.650 | $9.817\times10^{-1}$ | 3.602 |
| 0.0125 | $4.596\times10^{-3}$ | 3.372 | $2.603\times10^{-2}$ | 5.237 |
| 0.00625 | $1.915\times10^{-4}$ | 4.585 | $5.015\times10^{-4}$ | 5.698 |

**Table 3.** Comparison of the asset option with $(31c)$, $\varepsilon = 10^{-4}$.

| $S$ | $Benchmark\ Value$ [8] | | | | | |
|---|---|---|---|---|---|---|
| 100 | 6.9320 | | | | | |
| 110 | 4.1550 | | | | | |
| | $(\gamma_1, \gamma_2, \gamma_1, \gamma_2) = (2,3,4,5)$ | | | $(\gamma_1, \gamma_2, \gamma_1, \gamma_2) = (2,4,6,8)$ | | |
| | $h = 0.06$ | 0.03 | 0.01 | 0.06 | 0.03 | 0.01 |
| 100 | 6.9316 | 6.9323 | 6.9322 | 6.9310 | 6.9324 | 6.9322 |
| 110 | 4.1551 | 4.1552 | 4.1550 | 4.1543 | 4.1553 | 4.1550 |
| | $(\gamma_1, \gamma_2, \gamma_1, \gamma_2) = (3,4,5,6)$ | | | $(\gamma_1, \gamma_2, \gamma_1, \gamma_2) = (3,6,9,12)$ | | |
| 100 | 6.9294 | 6.9325 | 6.9322 | 6.9212 | 6.9326 | 6.9322 |
| 110 | 4.1524 | 4.1553 | 4.1550 | 4.1427 | 4.1554 | 4.1550 |

**Table 4.** Comparison of the optimal exercise boundary and its derivatives with $(34b)$, $CS-54$, $\varepsilon = 10^{-4}$.

| | $(\gamma_1, \gamma_2, \gamma_1, \gamma_2) = (2,3,4,5)$ | | |
|---|---|---|---|
| $h$ | 0.06 | 0.03 | 0.01 |
| $s_f(t)$ | 76.16 | 76.16 | 76.16 |
| $s_f'(t)$ | −4.59 | − 4.52 | − 4.51 |
| $s_f''(t)$ | 61.53 | 30.65 | 10.20 |
| | $(\gamma_1, \gamma_2, \gamma_1, \gamma_2) = (2,4,6,8)$ | | |
| $s_f(t)$ | 76.15 | 76.16 | 76.16 |
| $s_f'(t)$ | −4.63 | − 4.54 | − 4.51 |
| $s_f''(t)$ | 96.75 | 48.11 | 15.11 |
| | $(\gamma_1, \gamma_2, \gamma_1, \gamma_2) = (3,4,5,6)$ | | |
| $s_f(t)$ | 76.15 | 76.16 | 76.16 |
| $s_f'(t)$ | −4.68 | − 4.52 | − 4.51 |
| $s_f''(t)$ | 58.90 | 29.28 | 9.74 |
| | $(\gamma_1, \gamma_2, \gamma_1, \gamma_2) = (3,5,7,9)$ | | |
| $s_f(t)$ | 76.14 | 76.16 | 76.16 |
| $s_f'(t)$ | −4.47 | − 4.51 | − 4.51 |
| $s_f''(t)$ | 93.65 | 46.74 | 15.53 |
| | $(\gamma_1, \gamma_2, \gamma_1, \gamma_2) = (3,6,9,12)$ | | |
| $s_f(t)$ | 76.12 | 76.16 | 76.16 |
| $s_f'(t)$ | −4.48 | − 4.52 | − 4.51 |
| $s_f''(t)$ | 128.63 | 64.25 | 21.33 |

**Table 5a.** Comparing total CPU time(s) between SSPRK3 and 3(2) Bogacki-Shampine pairs with $(34c)$, $\varrho = 0.9$, $CS-54$, and $(\gamma_1, \gamma_2, \gamma_1, \gamma_2) = (2, 4, 6, 8)$. True value for $S = 90$ is 11.6976.

| | $SSPRK3$ | | | | | |
|---|---|---|---|---|---|---|
| | CPU time(s) | | | $S = 90$ | | |
| $h$ | $k = 4.0 \times 10^{-3}$ | $8.0 \times 10^{-4}$ | $4.0 \times 10^{-4}$ | $4.0 \times 10^{-3}$ | $8.0 \times 10^{-4}$ | $4.0 \times 10^{-4}$ |
| 0.02 | 17 | 43 | 85 | 11.6967 | 11.6976 | 11.6976 |
| | $3(2)\ Bogacki-Shampine\ pairs,\quad \varepsilon = 10^{-2}$ | | | | | |
| | $CPU\ time(s)$ | $S = 90$ | $Min.Time\ Step$ | $Ave.Time\ Step$ | $Max.Time\ Step$ | |
| 0.02 | 11 | 11.6976 | $4.9 \times 10^{-4}$ | $7.1 \times 10^{-3}$ | $1.6 \times 10^{-2}$ | |

**Table 5b.** Comparing total CPU time(s) with 3(2) Bogacki-Shampine pairs using, $CS-54, (\gamma_1, \gamma_2, \gamma_1, \gamma_2) = (2, 4,6, 8),\ \varepsilon = 10^{-2},\ h = 0.02,$ varying $\varrho$ and $(34c)$.

| $\varrho$ | $CPU\ time(s)$ | $S = 90$ |
|---|---|---|
| 0.20 | 8 | 11.6976 |
| 0.30 | 7 | 11.6976 |
| 0.40 | 10 | 11.6976 |
| 0.50 | 12 | 11.6976 |
| 0.55 | 11 | 11.6976 |
| 0.60 | 10 | 11.6976 |
| 0.65 | 11 | 11.6976 |
| 0.70 | 12 | 11.6976 |
| 0.75 | 12 | 11.6976 |
| 0.80 | 11 | 11.6976 |
| 0.85 | 12 | 11.6976 |

Furthermore, from Fig 2, we also observe that the numerical solutions of the first- and second-order derivatives of the optimal exercise are very smooth. Reflecting on the high convergence rate we obtained from the first-order derivative of the optimal exercise boundary, we can argue that our method is very efficient in approximating the optimal exercise boundary and its derivatives.

Lastly, but not least, the plot profile of the optimal time step for each time level was displayed in Fig. 3 and we observed the sensitivity of the optimal time step for each time level with respect to the interest rate, volatility, and the strike price. This is because the option value and its numerical approximation strongly depend on these parameters. By formulating how we select the time step adaptively for each time level, we are now interested in how the optimal time step for each time level changes across these important parameters when the latter is varied. We observed that the optimal time step selection is almost independent of the varying strike prices and interest rates but strongly dependent on the varying volatilities. If the other parameters are fixed as shown in Fig. $3c$, a decrease in volatility value results in large time step selection for each time level and vice versa. Without the implementation of adaptive time stepping, likely, the stability of our numerical scheme will strongly depend on the volatility parameter.

Like other RK-embedded pairs, we have implemented in our previous works, as one can easily observe in Fig. 3, a very small time step is required at the payoff and its neighborhood for any set of parameters chosen. This is one of the well-known features of the Runge-Kutta adaptive time integration methods that allow the selection of small or large time steps in regions where there is high variation, oscillation, and/or discontinuity or sufficient smoothness, respectively. It is much expected because of the observable discontinuity in this pricing model which occurs at the payoff.

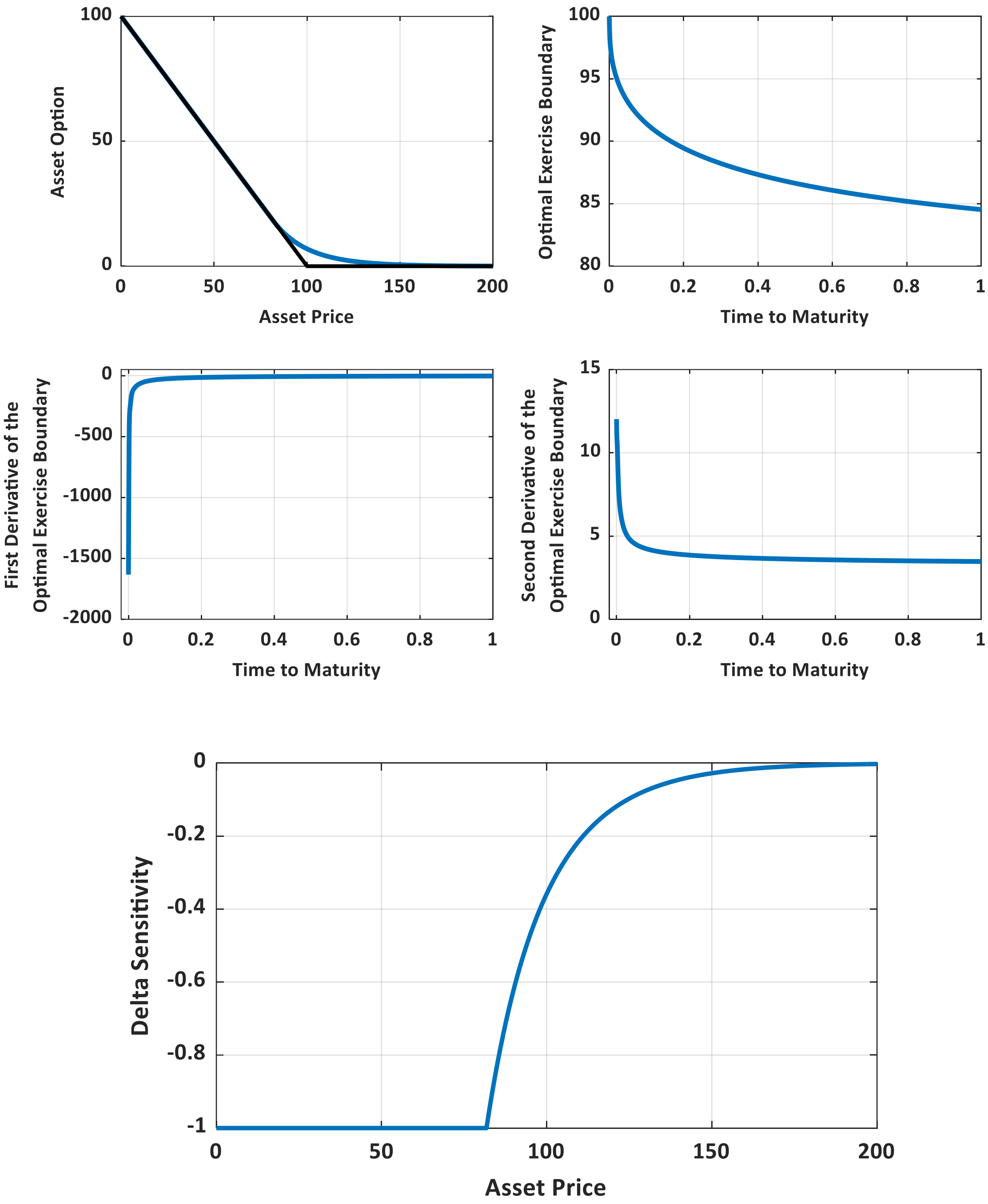

**Fig. 2.** Plot of asset option, delta sensitivity, optimal exercise boundary, and the derivatives of the optimal exercise boundary with a long time to maturity $(34c)$, $CS - 54$, $h = 0.01$, $\varepsilon = 10^{-4}$, and $(\gamma_1, \gamma_2, \gamma_1, \gamma_2) = (2{,}4{,}6{,}8)$.

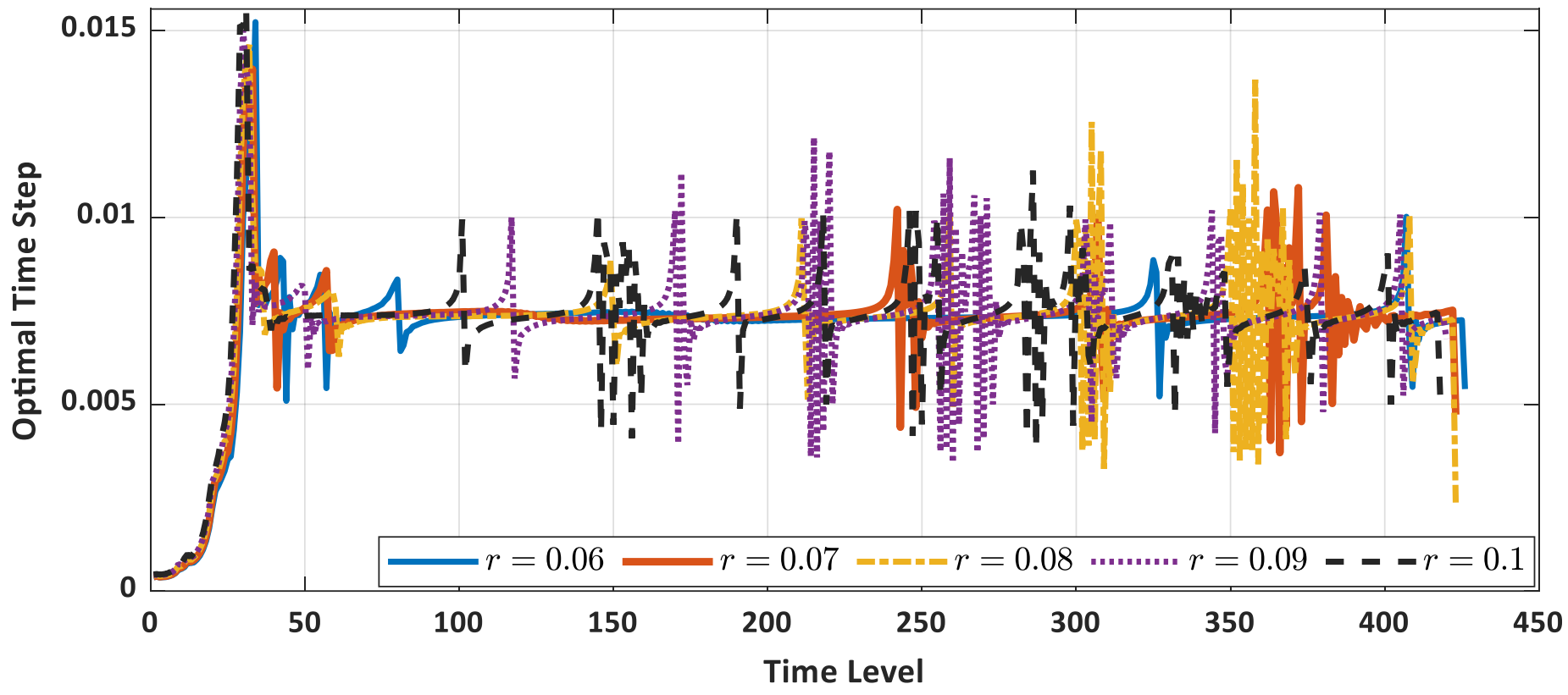

**Fig. 3a.** Profile of the optimal time step for each time level with $CS-54$, $h = 0.02$, $\varepsilon = 10^{-2}, (\gamma_1, \gamma_2, \gamma_1, \gamma_2) = (2,3,4,5)$, and 3(2) RK-Bogacki and Shampine embedded pairs. $K = 100$, $\sigma = 0.2, r = 0.06, 0.07, 0.08, 0.09, 0.1$.

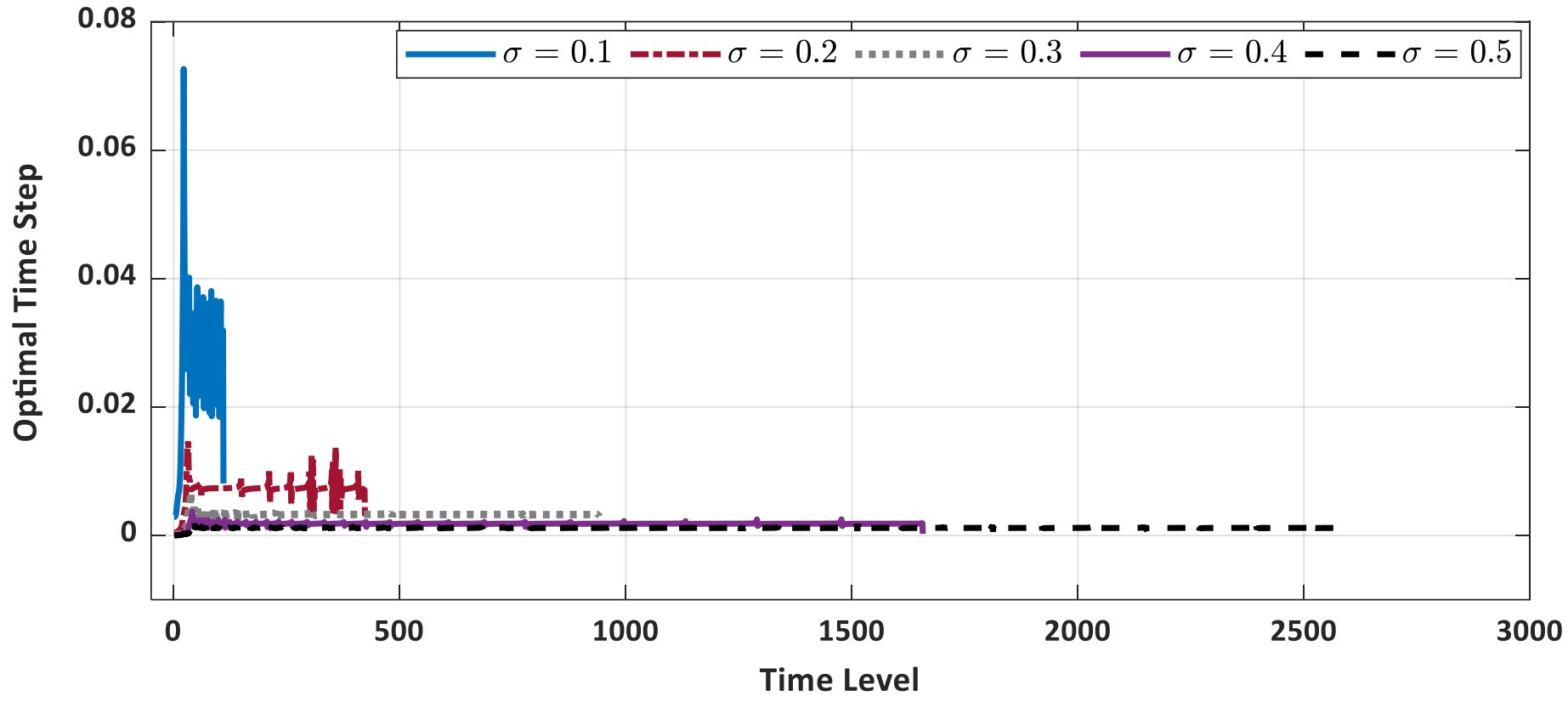

**Fig. 3b.** Profile of the optimal time step for each time level with $CS-54$, $h = 0.02$, $\varepsilon = 10^{-2}, (\gamma_1, \gamma_2, \gamma_1, \gamma_2) = (2,3,4,5)$ and 3(2) RK-Bogacki and Shampine embedded pairs. $K = 100$, $r = 0.08$, $\sigma = 0.1, 0.2, 0.3, 0.4, 0.5$.

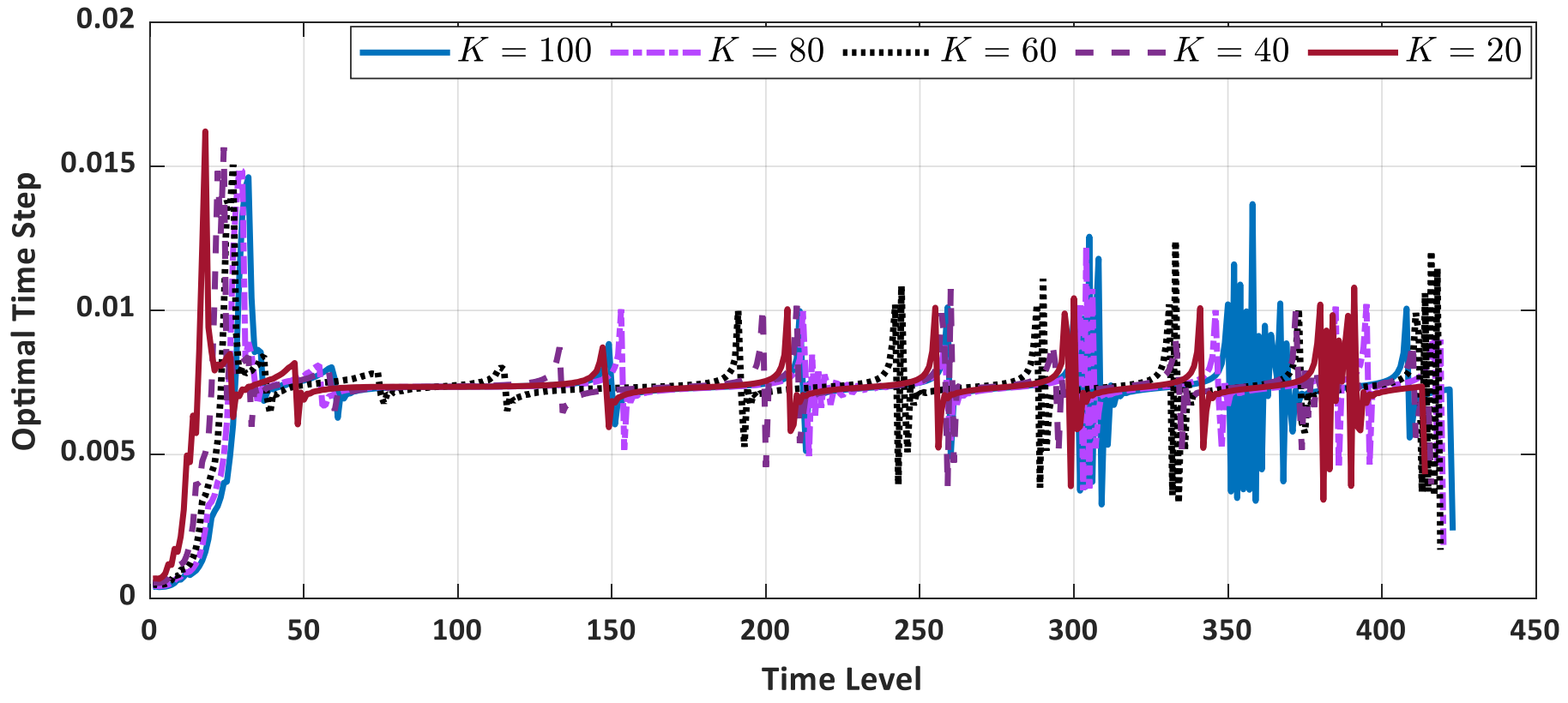

**Fig. 3c.** Profile of the optimal time step for each time level with $CS-54$, $h = 0.02$, $\varepsilon = 10^{-2}, (\gamma_1, \gamma_2, \gamma_1, \gamma_2) = (2,3,4,5)$ and 3(2) RK-Bogacki and Shampine embedded pairs. $r = 0.08$, $\sigma = 0.2$, and $K = 100, 80, 60, 40, 20$.

## 4. Conclusion

The front-fixing approach has shown its efficiency in locating, fixing, and precisely approximating the optimal exercise boundary when compared with the linear complementary problem and penalty method. However, this advantage is greatly hampered by the formulation of a nonlinear free-boundary problem with a discontinuous coefficient. This discontinuous coefficient in the model is associated with the convective term and involves the derivative of the optimal exercise boundary which is not continuous at the payoff. Hence, a non-smoothness is introduced which could deteriorate the accuracy of the American options.

A highly accurate numerical result with less computational cost can be achieved simultaneously when pricing the American options with the front-fixing approach if this non-smoothness is handled efficiently. Secondly, with an efficient implementation of high order numerical scheme, high order convergence rate can also be recovered from high order numerical scheme, and this entails obtaining more accurate numerical solutions with very coarse grids.

In this work, we aim at alleviating the challenges associated with solving American options using the front-fixing approach by first proposing a novel fifth-order staggered boundary scheme for deriving a high-order analytical approximation that improves the non-smoothness of the model. Furthermore, the precise values of the optimal exercise boundary, its derivatives, and the boundary values of the asset option and delta sensitivity are computed from this high-order analytical approximation. Coupled with an efficient sixth-order compact finite difference scheme, novel fifth- and sixth-order near-boundary schemes, and 3(2) RK-Bogacki and Shampine embedded pairs, highly numerical approximations of the interior and boundary values are obtained with a rather coarse grid and expected high-order convergence rate.

## Funding

The research is funded in part by an NSERC Discovery Grant.